\documentclass{SciPost}

\hypersetup{
    colorlinks,
    linkcolor={red!50!black},
    citecolor={blue!50!black},
    urlcolor={blue!80!black}
}

\usepackage{bm}
\usepackage{comment}
\usepackage{float} % added
\usepackage[bitstream-charter]{mathdesign}
\DeclareSymbolFont{usualmathcal}{OMS}{cmsy}{m}{n}
\DeclareSymbolFontAlphabet{\mathcal}{usualmathcal}

\fancypagestyle{SPstyle}{
\fancyhf{}
\lhead{\colorbox{scipostblue}{\bf \color{white} ~SciPost Physics Core }}
\rhead{{\bf \color{scipostdeepblue} ~Submission }}

\fancyfoot[C]{\textbf{\thepage}}
}

\begin{document}

\pagestyle{SPstyle}

\begin{center}{\Large \textbf{\color{scipostdeepblue}{
%%%%%%%%%% TODO: Write your article's title here
Energetic and Structural Properties of Two\hspace{+0.30mm}-\hspace{+0.05mm}Dimensional Trapped Mesoscopic Fermi Gases\\
%%%%%%%%%% END TODO: TITLE
}}}\end{center}

\begin{center}\textbf{
%%%%%%%%%% TODO: AUTHORS
% Write the author list here. 
% Use (full) first name (+ middle name initials) + surname format.
% Separate subsequent authors by a comma, omit comma and use "and" for the last author.
% Mark the corresponding author(s) with a superscript symbol in this order
% \star, \dagger, \ddagger, \circ, \S, \P, \parallel, ...
Emma~K.~Laird\textsuperscript{1,$\star$},
Brendan~C.~Mulkerin\textsuperscript{2},
Jia~Wang\textsuperscript{3}, and
Matthew~J.~Davis\textsuperscript{1}
%%%%%%%%%% END TODO: AUTHORS
}\end{center}

\begin{center}
%%%%%%%%%% TODO: AFFILIATIONS
% Write all affiliations here.
% Format: institute, city, country
{\bf 1} ARC Centre of Excellence in Future Low-Energy Electronics and Technologies, University of Queensland, Saint Lucia Queensland 4072, Australia
\\
{\bf 2} ARC Centre of Excellence in Future Low-Energy Electronics and Technologies, Monash University, Clayton Victoria 3800, Australia
\\
{\bf 3} Centre for Quantum Technology and Theory, Swinburne University of Technology, Hawthorn Victoria 3122, Australia
%%%%%%%%%% END TODO: AFFILIATIONS
%%%%%%%%%% TODO: EMAIL
% Provide email address of corresponding author(s)
\\[\baselineskip]
$\star$ \href{mailto:email1}{\small e.laird@uq.edu.au}%\,,\quad
%$\dagger$ \href{mailto:email2}{\small email2}
%%%%%%%%%% END TODO: EMAIL
\end{center}

\section*{\color{scipostdeepblue}{Abstract}}
\textbf{\boldmath{%
%%%%%%%%%% TODO: ABSTRACT
% Write your abstract here.
\raggedright We theoretically investigate equal-mass spin-balanced two-component Fermi gases in which pairs of atoms with opposite spins interact via a short-range isotropic model potential.  We probe the distinction between two-dimensional and quasi-two-dimensional harmonic confinement by tuning the effective range parameter within two-dimensional scattering theory.  Our approach, which yields numerically exact energetic and structural properties, combines a correlated Gaussian basis-set expansion with the stochastic variational method.  For systems containing up to six particles, we:  1) Present the ground- and excited-state energy spectra;  2) Study non-local correlations by analysing the one- and two-body density matrices, extracting from these the occupation numbers of the natural orbitals, the momentum distributions of atoms and pairs, and the molecular `condensate fraction';  3) Study local correlations by computing the radial and pair distribution functions.  This paper extends current theoretical knowledge on the properties of trapped few-fermion systems as realised in state-of-the-art cold-atom experiments.
%%%%%%%%%% END TODO: ABSTRACT
}}

\vspace{\baselineskip}

%%%%%%%%%% BLOCK: Copyright information
% This block will be filled during the proof stage, and finilized just before publication.
% It exists here only as a placeholder, and should not be modified by authors.
\noindent\textcolor{white!90!black}{%
\fbox{\parbox{0.975\linewidth}{%
\textcolor{white!40!black}{\begin{tabular}{lr}%
  \begin{minipage}{0.6\textwidth}%
    {\small Copyright attribution to authors. \newline
    This work is a submission to SciPost Physics Core. \newline
    License information to appear upon publication. \newline
    Publication information to appear upon publication.}
  \end{minipage} & \begin{minipage}{0.4\textwidth}
    {\small Received Date \newline Accepted Date \newline Published Date}%
  \end{minipage}
\end{tabular}}
}}
}
%%%%%%%%%% BLOCK: Copyright information

%%%%%%%%%% TODO: LINENO
% For convenience during refereeing we turn on line numbers:
% \linenumbers
% You should run LaTeX twice in order for the line numbers to appear.
%%%%%%%%%% END TODO: LINENO

%%%%%%%%%% TODO: TOC 
% Guideline: if your paper is longer that 6 pages, include a TOC
% To remove the TOC, simply cut the following block
\vspace{10pt}
\noindent\rule{\textwidth}{1pt}
\tableofcontents
\noindent\rule{\textwidth}{1pt}
% \vspace{10pt}
%%%%%%%%%% END TODO: TOC

%%%%%%%%% TODO: CONTENTS 
% Write your article contents here, starting from first \section.
% An example structure is given below.

\newcommand{\x}{\mathbf{x}}
\newcommand{\y}{\mathbf{y}}
\newcommand{\K}{\mathbf{K}}
\newcommand{\R}{\mathbf{R}}
\newcommand{\X}{\mathbf{X}}
\newcommand{\nn}{\nonumber}
\newcommand{\up}{\uparrow}
\newcommand{\down}{\downarrow}
\newcommand{\todo}[1]{{\color{blue} #1}}

\renewcommand{\k}{\mathbf{k}}
\renewcommand{\r}{\mathbf{r}}

\allowdisplaybreaks

% \raggedright

\section{Introduction}
\label{sec:Introduction}

Many-body quantum systems are generally intractable due to their vast complexity and numerous degrees of freedom.  A few of the simplest cases --- such as the Lieb--Liniger model of the one-dimensional Bose gas, or the one-dimensional Fermi--Hubbard model --- admit exact analytical solutions as a result of being integrable, but these are rare exceptions.  One prom-\linebreak ising strategy for understanding how many-body features emerge in more realistic settings is to probe the fundamental physics from the few-body limit.  Because the two-body system is\linebreak typically well characterised a ‘\hspace{+0.20mm}bottom-up\hspace{+0.15mm}’ approach can be conceived in which the number of particles is increased one by one, thereby introducing complexity in a controlled and stepwise fashion.  This treatment can sometimes reveal that mesoscopic observables converge surprisingly rapidly towards the predictions of many-body theories, once those predictions are rescaled to account for varying particle number~\cite{Wenz_2013,Zinner_2014,Grining_2015,Rammelmuller_2016,Levinsen_2017,Ran_2017,Schiulaz_2018,Bayha_2020,Holten_2022}.

An experimental bottom-up approach has been realised by the research group of Selim Jochim, who isolate a small number of ultracold fermionic atoms in a tightly focused optical microtrap. By superimposing this microtrap onto a larger reservoir of fermions and gradual-\linebreak ly lowering its depth, they can deterministically prepare a chosen number of particles in the ground state of a harmonic oscillator potential, close to zero temperature~\cite{Serwane_2011,Wenz_2013,Bayha_2020,Holten_2022}.  Applying\linebreak this method to two-component Fermi gases, the experiments have shown that in quasi-one-dimensional geometries a many-body Fermi sea can form with only four atoms~\cite{Wenz_2013}.  In quasi-two dimensions many-body `\hspace{+0.15mm}Cooper-like\hspace{+0.15mm}' pairing --- evidenced by a peak in the correlations between particles with opposing spins and momenta at the Fermi surface --- has been experimentally observed with as few as twelve atoms~\cite{Holten_2022}.

To better understand the latter experiment, in Ref.~\cite{Laird_2024} we theoretically investigated an increasing number of spin-balanced two-component fermions confined in a quasi-two-dimen-\linebreak sional harmonic trap.  Our numerical approach --- commonly known as the explicitly correlated Gaussian (ECG) method~\cite{Varga_&_Suzuki_1995,Varga_&_Suzuki_1996,Suzuki_&_Varga_1998,Review_ECG_Method} --- complemented a stochastic variational framework with the use of ECG basis functions~\cite{Boys_1960,Singer_1960}\hspace{+0.1mm}, allowing us to compute experimentally measurable observables with very high accuracy.  In particular, we calculated the lowest monopole excitation energies and ground-state opposite-spin pair correlations as functions of increasing attractive interaction strength~\cite{Laird_2024}.  The few-body physics was fully captured by applying two-dimensional scattering theory~\cite{Verhaar_1984,Adhikari_1986_A,Adhikari_1986_B} to a finite-range Gaussian interaction potential and tuning the effective range to model realistic quasi-two-dimensional scattering~\cite{Levinsen_2013,Kirk_2017,Hu_2019,Yin_2020}.  For gases comprising up to six equal-mass fermions, we found that time-reversed pairing in the ground state was dominant at momenta significantly below the Fermi momentum~\cite{Laird_2024}.  Together with experimental findings~\cite{Holten_2022}\hspace{+0.1mm}, this suggested that the Fermi sea --- which, beneath the Fermi surface, Pauli-blocks the superposition of momenta required to form a paired state --- must emerge in the transition from six to twelve particles.

Here, we apply the ECG method to the same Fermi gases to obtain new energy spectra and ground-state structural properties, which are crucial for their theoretical characterisation and thereby further advance our understanding of fermionic few-body systems.  This paper is organised as follows:  In Section~\ref{sec:Model} we outline our model and the underlying two-body scattering theory.  Section~\ref{sec:Results} details our results: In Subsection~\ref{sec:Energy_Spectra} we generate the energy spectra of the ground state and low-lying excited-state manifolds for gases containing two, four, and six particles.  We quantify non-local correlations between the trapped fermions  by analysing the one- and two-body density matrices in Subsection~\ref{sec:Density_Matrices_and_Occupation_Numbers}. In Subsection~\ref{sec:Momentum_Distributions} we analytically Fourier transform the density matrices to extract the momentum distributions of individual atoms and opposite-spin pairs. To quantify local correlations in the Fermi gases we examine the radial and pair distribution functions in Subsection~\ref{sec:Radial_and_Pair_Distribution_Functions}.  In Subsection~\ref{sec:Finite_Effective_Range} we elucidate the effect of the trap aspect ratio --- i.e., effective range --- on the energetic and structural properties men-\linebreak tioned above.  We conclude and discuss the relative merits of our approach in Section~\ref{sec:Conclusion}.  Our work is strongly inspired by earlier, similar studies of trapped few-fermion systems subject to three-dimensional harmonic confinement --- particularly Ref.~\cite{Blume_2011}\hspace{+0.1mm}, as well as Refs.~\cite{von_Stecher_2008,Daily_2010,Bradly_2014,Yin_2015}. These publications, in turn, are partly motivated by research on bosonic $^{4}$He and fermionic $^{3}$He droplets~\cite{Lewart_1988} which, although much denser than ultracold atomic gases, can be described using the same theoretical framework.

\section{Model}
\label{sec:Model}

The two-component Fermi gases considered in our study consist of equal-mass atoms with balanced spin populations, such that $N=N_\up\hspace{-0.2mm}+\hspace{+0.2mm}N_\down$ and $N_\up=N_\down=N/2$\hspace{+0.2mm}, where $N_\up$ and $N_\down$ denote the number of `\hspace{+0.2mm}spin-up\hspace{+0.2mm}' and `\hspace{+0.2mm}spin-down\hspace{+0.1mm}' fermions, respectively.  Each gas is confined in an isotropic two-dimensional (2D) harmonic trap and in the non-interacting ground state only the first two harmonic oscillator shells are occupied --- corresponding to particle numbers,  $N_\up\hspace{-0.2mm}+\hspace{+0.2mm}N_\down=1+1$, $2+2\hspace{+0.3mm}$, and $3+3$.  Our work is inspired by recent experiments conducted in the group of Selim Jochim~\cite{Bayha_2020,Holten_2022}\hspace{+0.1mm}, in which the harmonically trapped ground state of a small number of fermionic $^{6}\mathrm{Li}$ atoms --- ranging from 20 down to just 2 --- can be prepared with very high fidelity.

The effective low-energy Hamiltonian reads as follows:
\begin{align}
\label{eq:Hamiltonian}
\mathcal{H}=
\sum_{i\,=\,1}^N\left[-\frac{\hbar^2}{2m}\nabla^{\hspace{+0.2mm}2}_{\hspace{-0.6mm}\r_i}+{V}_{\mathrm{ext}}(|\r_i|)\right]+\sum_{i\,<\,j}^N{V}_{\mathrm{int}}(|\r_i-\r_j|)\;,
\end{align}
where $m$ is the atomic mass and $\r_i$ is the 2D position vector of the $i^{t\hspace{-0.2mm}h}$ atom measured from the centre of the trap.  The first term corresponds to the kinetic energy, the second term to the external confinement,
\begin{align}
\label{eq:trapping_potential}
{V}_{\mathrm{ext}}(|\r_i|)=\frac{m\omega_{\hspace{-0.2mm}r}^2}{2}\hspace{+0.4mm}r_{\hspace{-0.2mm}i}^{\hspace{+0.1mm}2}\hspace{-0.4mm}\,,\quad{r}_i\equiv|\r_i|\hspace{+0.1mm}\;,
\end{align}
where $\omega_r$ is the radial harmonic trapping frequency, and the third term to short-range pairwise interactions.  Note that Pauli exclusion ensures identical fermions do not interact.  The interactions between distinguishable fermions are described using a finite-range Gaussian potential, parameterised by a width $r_0$ $(>0)$ and a depth $V_0$ $(<0)\hspace{0.0mm}$:
\begin{align}
\label{eq:interaction_potential}
V_{\mathrm{int}}(|\r\hspace{+0.2mm}|)=V_0\hspace{+0.5mm}\mathrm{exp}\hspace{-0.3mm}\left(-\hspace{+0.2mm}\frac{r^2}{2r_{\hspace{-0.2mm}0}^2}\right)-\hspace{+0.1mm}V_0\hspace{+0.3mm}\frac{r}{l_{r}}\hspace{+0.5mm}\mathrm{exp}\hspace{-0.3mm}\left[-\hspace{+0.2mm}\frac{r^2}{2\hspace{+0.1mm}(2r_0)^2}\right]\hspace{-0.2mm}\,.
\end{align}
Here, $l_{r}=\sqrt{\hbar/(m\omega_r)}$ is the radial harmonic oscillator length scale in the plane.  This potential has previously been employed to model the breathing modes~\cite{Yin_2020} and time-reversed pair correlations~\cite{Laird_2024} of a few interacting fermions in a 2D harmonic trap.  In the non-interacting limit of $V_0=0$ the eigenvalues of the Hamiltonian~\eqref{eq:Hamiltonian} are $\varepsilon^{(0)}_{nm}=(2n+|m|+1)\,\hbar\omega_r$\hspace{+0.1mm}, where $n=0,\,1,\,2\hspace{+0.20mm},\,\dots$ is the principal quantum number and $m=0,\,\pm1,\,\pm2\hspace{+0.20mm},\,\dots$ is the quantum number for orbital angular momentum.

The values of $r_0$ and $V_0$ can be adjusted to generate potentials with different $s$\hspace{-0.1mm}-wave scattering properties in 2D free space~\cite{Sakurai_2010}. We solve the $s$\hspace{-0.1mm}-wave radial Schr\"{o}dinger equation for the relative motion of two elastically scattering atoms, matching the logarithmic derivatives of the wave functions inside and outside the range of the interaction potential~\eqref{eq:interaction_potential} to obtain the scattering phase shift $\delta(k)$.  By subsequently fitting the phase shift to the known form~\cite{Verhaar_1984,Adhikari_1986_A,Adhikari_1986_B} of its low-energy expansion in two dimensions,
\begin{align}
\label{eq:phase_shift_expansion}
\mathrm{cot}\hspace{+0.3mm}[\hspace{+0.1mm}\delta(k)]=\frac{2}{\pi}\left[\hspace{+0.1mm}\gamma+\mathrm{ln}\left(\frac{k{a}_\mathrm{2D}}{2}\right)\hspace{-0.3mm}\right]+\frac1\pi{k}^2r_\mathrm{2D}+\dots\;,
\end{align}
we ascertain both the $s$\hspace{-0.1mm}-wave scattering length $a_\mathrm{2D}$ and effective range $r_\mathrm{2D}$\hspace{+0.1mm}.\footnote{\hspace{+0.1mm}Note that the exact definitions of the 2D scattering length  and  effective range are not fixed in the literature.  Our definition of $r_\mathrm{2D}$ has units of squared length, consistent with Refs.~\cite{Yin_2020,Laird_2024}.}  Above, $k\equiv|\hspace{+0.20mm}\k\hspace{+0.35mm}|$ is the magnitude of the relative wave vector between the atoms in the plane and $\gamma\simeq0.577216$ is Euler's constant.  Importantly, the low-energy physics does not depend on the short-range details of the true interaction potential and is, instead, universally determined by $a_\mathrm{2D}$ and $r_\mathrm{2D}$\hspace{+0.1mm}.\linebreak  In all our calculations, we therefore choose Gaussian widths small enough ($r_0\lesssim0.1l_r$) that higher order terms in the expansion~\eqref{eq:phase_shift_expansion} are negligible in the energy range of interest.  In two\linebreak dimensions a two-body bound state always exists — even for arbitrarily weak attractive interactions — since the scattering amplitude obtained by the analytic continuation of Eq.~\eqref{eq:phase_shift_expansion} to negative energies always exhibits a pole.  In the zero effective range limit, the corresponding binding energy $\varepsilon_b$ is related to the 2D scattering length via $\varepsilon_b = 4\hspace{+0.3mm}\hbar^{\hspace{+0.2mm}2}e^{-\hspace{+0.2mm}2\hspace{+0.2mm}\gamma}\hspace{-0.3mm}/\hspace{+0.3mm}(m \hspace{+0.1mm}a_\mathrm{2D}^2\hspace{+0.2mm})$\hspace{+0.1mm}. For finite $r_\mathrm{2D}$ this relationship must be determined numerically from the phase shift expansion; however, $\varepsilon_b$ still serves as a monotonic proxy for interaction strength~\cite{Levinsen_&_Parish_2015}.

The scattering length is always positive ($a_\mathrm{2D}>0$) because it enters as the argument of the logarithm in Eq.~\eqref{eq:phase_shift_expansion} and the phase shift must remain real at low energies.  In the many-body limit as $a_\mathrm{2D}$ increases the two-component Fermi gas undergoes a crossover from a Bose--Ein-\linebreak stein condensate (BEC) of tightly bound diatomic molecules to a Bardeen--Cooper--Schrieffer (BCS) superfluid of long-range Cooper pairs~\cite{Levinsen_&_Parish_2015,Zwerger_2012}.  However, unlike in three dimensions, there is no unitary limit where the interaction strength diverges and becomes scale invariant.  Rather, the strongly interacting regime emerges around the point $\mathrm{ln}\hspace{+0.2mm}(k_Fa_\mathrm{2D})=0$\hspace{+0.1mm}, where the Fermi momentum $k_F$\hspace{-0.1mm} determines the average interparticle spacing~\cite{Levinsen_&_Parish_2015,Zwerger_2012}.  In the few-body limit this spacing becomes ill-defined due to large fluctuations, making the regime of strong interactions more difficult to characterise for only a small number of atoms.

A two-dimensional geometry is experimentally realised by applying a strong harmonic confinement along the axial direction~\cite{Bayha_2020,Holten_2022}\hspace{+0.1mm}, characterised by an angular frequency $\omega_z$ and corresponding length scale $l_{z}=\sqrt{\hbar/(m\omega_z)}$\hspace{+0.1mm}.  However, in reality, the gas extends a small but finite distance perpendicular to the plane.  At low energy, when $l_{z}$ is small (such that $k\hspace{+0.1mm}l_z\ll1$) but still much larger than the van der Waals range of the interactions, the two-body scattering of distinguishable fermions can be mapped to a purely 2D scattering amplitude with an effective range given by~\cite{Levinsen_2013,Kirk_2017,Hu_2019,Yin_2020}
\begin{align}
\label{eq:effective_range}
{r}_\mathrm{2D}&=-\hspace{+0.2mm}l_z^2\,\hspace{+0.2mm}\mathrm{ln}\hspace{+0.2mm}(2)\;.
\end{align}
As a result, the effect of a \textit{quasi}-2D geometry on the scattering can be mimicked and probed by attributing a finite, negative value to the effective range in the 2D model, Eqs.~\eqref{eq:Hamiltonian}\hspace{+0.2mm}--\hspace{+0.2mm}\eqref{eq:effective_range}.  The effective range can be tuned through a wide range of negative values near a shape resonance~\cite{Braaten_2006,Yin_2020} which arises due to the structure of the model potential.  Virtual bound states are supported in the attractive well associated with the first term of Eq.~\eqref{eq:interaction_potential}, and these can couple to free-space scattering states through the small repulsive barrier created by the second term.  We restrict our calculations to the regime where this potential supports a single two-body $s$\hspace{-0.1mm}-wave bound state in two-dimensional free space~\cite{Yin_2020,Laird_2024}.  In Subsections~\ref{sec:Energy_Spectra}--\hspace{+0.3mm}\ref{sec:Radial_and_Pair_Distribution_Functions} we fix the effective range to very nearly zero, $r_\mathrm{2D}/l_r^2=-\hspace{+0.2mm}0.001\approx0$, in order to determine the energetic and structural properties of the Fermi gases very close to the strictly 2D limit, which is of fundamental interest.  Increasing $|r_\mathrm{2D}|$ --- while remaining within the regime of the mapping in Eq.~\eqref{eq:effective_range} --- leads to small quantitative shifts in these results but, most of the time, leaves them qualitatively unchanged.  In Subsection~\ref{sec:Finite_Effective_Range} we show how our results are modified for $r_\mathrm{2D}/l_r^2=-\hspace{+0.2mm}0.2$ which was the largest negative value considered in Ref.~\cite{Laird_2024}.

\section{Results and Discussion}
\label{sec:Results}

To numerically solve the time-independent Schr\"{o}dinger equation for the Hamiltonian~\eqref{eq:Hamiltonian} we employ the explicitly correlated Gaussian method discussed in detail in our earlier publication~\cite{Laird_2024} (see Appendix~A therein).  Other works which also apply this technique to study ul-\linebreak tracold two-component fermions include Refs.~\cite{von_Stecher_2008,Daily_2010,Blume_2011,Bradly_2014,Yin_2015,Yin_2020}.  Our calculations are parameterised in terms of the two-body binding energy $\varepsilon_b\geq0$ and the effective range $r_\mathrm{2D}$\hspace{+0.1mm}.  Although $\varepsilon_b$ was\linebreak introduced in Section~\ref{sec:Model} in the context of free-space two-body scattering, it can additionally be defined in the presence of the harmonic trap. The two definitions coincide in the weak con-\linebreak finement limit and in both cases $\varepsilon_b$ remains a monotonic function of the underlying scattering parameters, $a_\mathrm{2D}$ and $r_\mathrm{2D}$\hspace{+0.1mm}.  In practice, we obtain $\varepsilon_b$ by using the ECG method to calculate the relative ground-state energy $\varepsilon_\mathrm{rel}$ for $1+1$ fermions at given $r_0$ and $V_0$\hspace{+0.2mm}, which specify the interaction potential of Eq.~\eqref{eq:Hamiltonian}.  The total ground-state energy in the trap takes the form $\varepsilon=\linebreak\varepsilon_\mathrm{com}\hspace{-0.1mm}+\hspace{+0.2mm}\varepsilon_\mathrm{rel}=2\hspace{+0.2mm}\hbar\omega_r-\varepsilon_b$\hspace{+0.1mm}, and since there are no centre-of-mass excitations in the ground state $\varepsilon_\mathrm{com}=\hbar\omega_r$\hspace{+0.1mm}, we can immediately find $\varepsilon_b$\hspace{0mm}.

\subsection{Energy Spectra}
\label{sec:Energy_Spectra}

\begin{figure}[H]
\begin{center}
\hspace{0mm}
\includegraphics[trim = 0mm 0mm 0mm 0mm, clip, scale = 0.68]{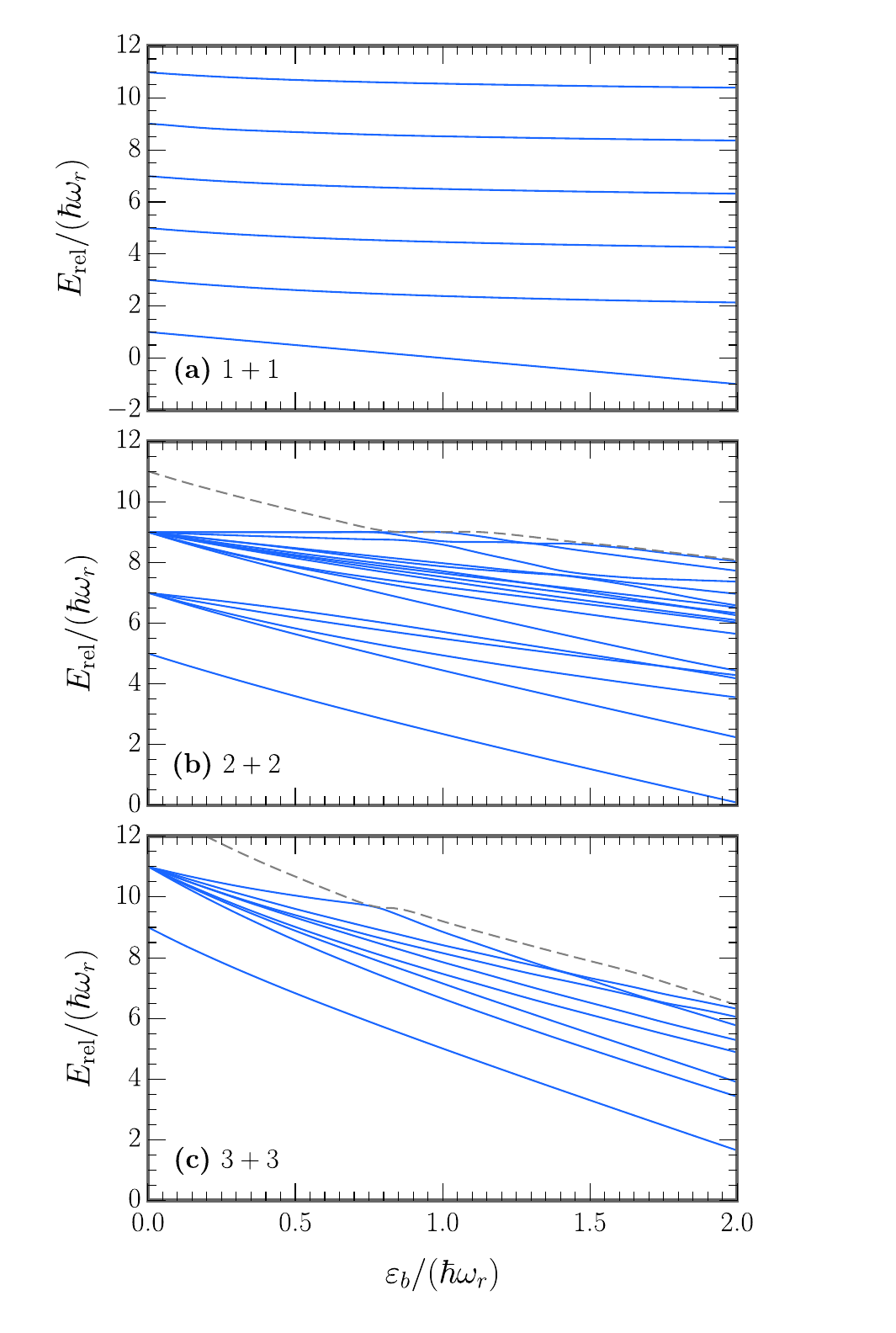}
\caption{The monopole energy spectrum of (a) $1+1$, (b) $2+2$\hspace{+0.35mm}, and (c) $3+3$ fer-\linebreak mions at very nearly zero effective, $r_\mathrm{2D}/l_r^2=-\hspace{+0.2mm}0.001\approx0$.  $E_\mathrm{rel}$ is the total relative energy and $\varepsilon_b$ is the two-body binding energy.  In panels (b) and (c) the grey dashed line indicates the energy of the first state of the next (unshown) manifold.}
\label{fig:Energy_Spectra}
\end{center}
\end{figure}

In two dimensions the exact energy spectrum for $1+1$ fermions was analytically calculated\linebreak by Busch \emph{et al.}~in 1998~\cite{Busch_1998}.  Their approach involved modelling the interaction with a regularised Dirac delta distribution, expanding the relative wave function in the harmonic oscillator basis, and then using standard integral representations to evaluate the Schr\"{o}dinger equation.  In 2010 Liu \emph{et al.}~numerically computed the exact energy spectrum for $2+1$ ferm-\linebreak ions by extending the approach of Efimov~\cite{Efimov_1971} to the two-dimensional trapped case and applying the Bethe--Peierls boundary condition~\cite{Liu_2010}.  In this subsection we obtain numerically exact energy spectra for $1+1$, $2+2$\hspace{+0.35mm}, and $3+3$ fermions at very nearly zero effective range, $r_\mathrm{2D}/l_r^2=-\hspace{+0.2mm}0.001\approx0$.  After separating off the centre-of-mass degree of freedom, we expand the eigenstates of the relative Hamiltonian in terms of explicitly correlated Gaussian basis\linebreak functions~\cite{Laird_2024,Varga_&_Suzuki_1995,Varga_&_Suzuki_1996,Suzuki_&_Varga_1998,Review_ECG_Method}.  These basis functions depend on a set of non-linear variational parameters (the Gaussian widths) which are optimised by energy minimisation. In Fig.~\ref{fig:Energy_Spectra} we plot the resultant energies as functions of the two-body binding energy $\varepsilon_b$\hspace{0mm}.

The non-interacting ground state at $\varepsilon_b=0$ can assume one of two configurations depending on the total number of particles $N$\hspace{-0.1mm}:  either all of the degenerate single-particle states of the highest energy level of the 2D harmonic oscillator are filled (`\hspace{+0.1mm}closed shell\hspace{+0.2mm}'), or some of the degenerate states remain empty (`\hspace{+0.1mm}open shell\hspace{+0.2mm}'). The $1+1$ and $3+3$ systems both feature a closed-shell ground state that is non-degenerate, whereas the $2+2$ ground state is open-\linebreak shell. We restrict our analysis to ground states characterised by zero total orbital angular momentum. For the $2+2$ system this means that the two highest energy opposite-spin fermions reside in different degenerate single-particle states.  Since the Hamiltonian is rotationally sym-\newpage

\vspace{0.2mm}

\noindent metric, only monopole excitations between states with the same (i.e., zero) total angular momentum occur.  (The $m$ quantum numbers for all atoms sum to zero in both the ground and excited states.)  We can see in Fig.~\ref{fig:Energy_Spectra} that for all three atom numbers at $\varepsilon_b=0$, all monopole excitations have an energy of $2\hspace{+0.2mm}\hbar\omega_r$. This can be attributed either to exciting a single particle up two harmonic oscillator shells, or to exciting a time-reversed pair of particles $(n,\,m,\,\up)$ and $(n,\,-\hspace{+0.2mm}m,\,\down)$ up one shell each.

Our result for $1+1$ fermions in Fig.~\ref{fig:Energy_Spectra}(a) agrees with the `\hspace{+0.2mm}Busch spectrum\hspace{+0.1mm}'~\cite{Busch_1998} for the considered range of binding energies, $0\leq\varepsilon_b\leq2\hspace{+0.2mm}\hbar\omega_r$. As evident in Fig.~2 of Ref.~\cite{Laird_2024}\hspace{+0.1mm}, this range is sufficient to capture the non-monotonic dependence on $\varepsilon_b$ of the lowest monopole excitation of $3+3$ fermions [Fig.~\ref{fig:Energy_Spectra}(c)] --- a feature which is driven by coherent pair correlations~\cite{Bjerlin_2016}.  Larger basis sizes are required for the ECG method to converge at higher binding energies, $\varepsilon_b>2\hspace{+0.2mm}\hbar\omega_r$\hspace{+0.1mm}, where the tight composite bosonic wave functions become difficult to represent numerically~\cite{Laird_2024,Blume_2011}.  Currently, convergence  cannot be achieved in this regime for six atoms, although it may be possible for four (and is certainly possible for two).  It is additionally challenging to solve for more than six particles at \textit{any} binding energy due to the factorial growth (with $N$) in the number of permutations of identical fermions  required to antisymmetrise the full wave function~\cite{Laird_2024,Blume_2011}. The spectra in Fig.~\ref{fig:Energy_Spectra} for increasing $N$ are qualitatively similar, but increasingly complex due to the existence of higher degeneracies in the non-interacting limit.  For $1+1$ fermions [Fig.~\ref{fig:Energy_Spectra}(a)] we choose to show the six lowest energy states, while for $2+2$ fermions [Fig.~\ref{fig:Energy_Spectra}(b)] we choose to show the ground state and the first- and second-excited-state manifolds.  For $3+3$ fermions [Fig.~\ref{fig:Energy_Spectra}(c)] we display the ground state and the first-excited-state manifold which, in this case, is the largest number of states that can be computed to numerical convergence within a reasonable time frame (on the order of months).

\subsection{Density Matrices and Occupation Numbers}
\label{sec:Density_Matrices_and_Occupation_Numbers}

\subsubsection{One-Body Density Matrix}

In the first-quantised position representation the one-body density matrix for the spin-$\up$ particles is given by
\begin{align}
\label{eq:rho-up}
&\rho_\up(\r,\,\r')=\left[\hspace{+0.20mm}\int\!\cdots\!\int{d}\r_1^\up{d}\r_2^\down\cdots{d}\r_{N-1}^\up{d}\r_N^\down\left|\Psi(\r_1^\up,\,\r_2^\down,\,\cdots,\,\r_{N-1}^\up,\,\r_N^\down)\right|^2\hspace{+0.10mm}\right]^{\hspace{+0.25mm}-1}\hspace{-3.1mm}\times\nn\\
&\hspace{-2mm}\int\!\cdots\!\int{d}\r_2^\down\hspace{+0.30mm}{d}\r_3^\up\hspace{+0.20mm}{d}\r_4^\down\cdots{d}\r_{N-1}^\up{d}\r_N^\down\Psi(\r,\,\r_2^\down,\,\r_3^\up,\,\r_4^\down,\,\cdots,\,\r_{N-1}^\up,\,\r_N^\down)\hspace{+0.20mm}\Psi^*(\r'\hspace{-0.5mm},\,\r_2^\down,\,\r_3^\up,\,\r_4^\down,\,\cdots,\,\r_{N-1}^\up,\,\r_N^\down)\;,
\end{align}
where $\Psi$ is the total $N\hspace{-0.3mm}$-body wave function and all integrals are two-dimensional (\hspace{+0.1mm}$d\r\equiv{d}^{2}\hspace{+0.1mm}\r$\hspace{+0.2mm}).\linebreak  The first line above is a normalisation constant; in the second line the density $\Psi\Psi^*$ is integrated over all co-ordinates except those of a single spin-$\up$ atom.  The matrix elements of Eq.~\eqref{eq:rho-up} in the explicitly correlated Gaussian basis were derived in our earlier work --- refer to Appendices A, C, and D of Ref.~\cite{Laird_2024} --- and we quote the final result below for ease of reference:
\begin{align}
\label{eq:final_rho-up}
[\hspace{+0.2mm}\rho_\up(\r,\,\r')]_{\mathbb{A}\hspace{+0.1mm}\mathbb{A}'}\equiv\langle\phi_{\mathbb{A}}|\hspace{+0.4mm}\rho_\up(\r,\,\r')\hspace{+0.4mm}|\phi_{\mathbb{A}'}\rangle=\hspace{+0.2mm}c_1\,\hspace{-0.4mm}\mathrm{exp}\hspace{+0.1mm}\!\left\{-\hspace{+0.2mm}\frac12\hspace{+0.25mm}\Big[c\hspace{+0.1mm}\r^{\hspace{+0.3mm}2}+c'(\r'\hspace{+0.1mm})^{\hspace{+0.1mm}2}-a\hspace{+0.1mm}\r^T\hspace{-0.3mm}\r'\Big]\hspace{0.0mm}\right\}\hspace{-0.1mm},
\end{align}
which contains the following scalars:
\begin{subequations}
\begin{align}
\label{eq:c1}
&c_1=\frac{(2\pi)^{\hspace{0mm}N\hspace{-0.1mm}-1}}{\mathrm{det\hspace{+0.2mm}[\hspace{+0.2mm}\mathbb{B}+\mathbb{B}'\hspace{+0.2mm}]}}\hspace{+0.4mm}\,,\\
\label{eq:c}
&c=b_1-\mathbf{b}^T\hspace{-0.1mm}\mathbb{C}\hspace{+0.2mm}\mathbf{b}\,\hspace{+0.4mm},\\
\label{eq:c'}
&c\hspace{+0.2mm}'=b_1'-(\mathbf{b}')^T\hspace{-0.1mm}\mathbb{C}\hspace{+0.2mm}\mathbf{b}'\,\hspace{-0.5mm},\\
\label{eq:a}
&a=\mathbf{b}^T\hspace{-0.1mm}\mathbb{C}\hspace{+0.2mm}\mathbf{b}'+(\mathbf{b}')^T\hspace{-0.1mm}\mathbb{C}\hspace{+0.2mm}\mathbf{b}\,\hspace{+0.4mm}.
\end{align}
\end{subequations}
Here, $b_1=(\mathbb{U}^{\hspace{+0.2mm}T}\!\hspace{-0.25mm}\mathbb{A}\mathbb{U})_{11}$ is also a scalar, $\mathbf{b}=((\mathbb{U}^{\hspace{+0.2mm}T}\!\hspace{-0.25mm}\mathbb{A}\mathbb{U})_{12}\hspace{+0.3mm},\,\dots,\,(\mathbb{U}^{\hspace{+0.2mm}T}\!\hspace{-0.25mm}\mathbb{A}\mathbb{U})_{1\hspace{-0.1mm}N})\hspace{+0.3mm}$ is an $(N\hspace{-0.3mm}-1)$-dimension-\linebreak al vector, $\mathbb{B}$ is an $(N-1)\hspace{-0.25mm}\times(N-1)$-dimensional matrix given by $\mathbb{U}^{\hspace{+0.2mm}T}\!\hspace{-0.25mm}\mathbb{A}\mathbb{U}$ with the first row and column removed, and $\mathbb{C}=(\mathbb{B}+\mathbb{B}')^{-1}$.  The $N\hspace{-0.4mm}\times{N}$ transformation matrix  $\mathbb{U}$ ($\x=\mathbb{U}\hspace{+0.2mm}\y$) converts the single-particle co-ordinates $\y$ into relative and centre-of-mass generalised Jacobi co-ordin-\linebreak ates $\x$ (where $\x$ and $\y$ are vectors of vectors).  The $N\hspace{-0.4mm}\times{N}$ correlation matrix $\mathbb{A}$ comprises non-linear variational parameters (the Gaussian widths) which are optimised stochastically.  Each ECG basis function $|\phi_{\mathbb{A}}\hspace{+0.2mm}\rangle$ is numerically represented by a unique $\mathbb{A}$ matrix~\cite{Laird_2024}.

Equation~\eqref{eq:rho-up} can be expanded over a complete set of basis functions --- the natural orbitals $\chi_{nm}(\r)$ --- where the expansion coefficients correspond to the occupation numbers $\mathcal{N}_{nm}$ of those orbitals:
\begin{align}
\label{eq:1BDM_decomposition}
\rho_\up(\r,\,\r')=\sum_{nm}\mathcal{N}_{nm}\,\hspace{0mm}\chi_{nm}^*(\r)\,\chi_{nm}(\r')\;.
\end{align}
These components are normalised as follows:
\begin{subequations}
\begin{align}
\int\!d\r\,\chi_{nm}^*(\r)\,\chi_{n'\hspace{-0.15mm}m'}(\r)&=\delta_{nn'}\hspace{+0.2mm}\delta_{mm'}\;,\\
\label{eq:1BDM_occ-num_norm-cond}
\sum_{nm}\mathcal{N}_{nm}&=1\,\hspace{0mm},
\end{align}
\end{subequations}
where $\{n,\,m\}$ are the harmonic oscillator quantum numbers defined below Eq.~\eqref{eq:interaction_potential}, and where the asterisk denotes complex conjugation (although in our specific case the natural orbitals are real).  Since, in practice, it is not feasible to decompose the four-dimensional object $\rho_\up(\r,\,\r')$ directly as written in Eq.~\eqref{eq:1BDM_decomposition}, we first reduce the number of degrees of freedom by defining partial-wave projections (see Ref.~\cite{Blume_2011} for an analogous treatment in three dimensions):
\begin{align}
\label{eq:partial-wave_projections}
\rho_\up^{m}(r,\,r')=\frac{1}{2\pi}\int_0^{2\pi}\!\int_0^{2\pi}\!d\hspace{-0.3mm}\theta\hspace{+0.3mm}{d}\hspace{-0.3mm}\theta'\hspace{-0.1mm}e^{-im\theta}\hspace{-0.3mm}\rho_\up(\r,\,\r')\,\hspace{+0.1mm}e^{im\theta'}\hspace{-0.5mm},
\end{align}
with $\theta^{(}$$'$$^{)}$ denoting the angle associated with the vector $\r^{\hspace{+0.4mm}(}$$'$$^{)}$ and $r^{\hspace{+0.1mm}(}$$'$$^{)}\hspace{-0.1mm}\equiv|\r^{\hspace{+0.4mm}(}$$'$$^{)}|$.  The explicitly correlated Gaussian matrix elements of Eq.~\eqref{eq:partial-wave_projections} are
\begin{align}
\label{eq:projected_1BDM}
[\hspace{+0.2mm}\rho_\up^m(r,\,r')]_{\mathbb{A}\hspace{+0.1mm}\mathbb{A}'}\equiv\langle\phi_{\mathbb{A}}|\hspace{+0.4mm}\rho_\up^m\hspace{+0.4mm}|\phi_{\mathbb{A}'}\rangle=2\pi%\,\hspace{+0.1mm}(\mathbb{O}_{\mathbb{A}\mathbb{A}'})^{-1}\hspace{-0.3mm}
\,c_1\,\mathcal{I}_m\!\hspace{+0.1mm}\left(\frac{arr'}{2}\right)\hspace{+0.1mm}\mathrm{exp}\hspace{+0.1mm}\!\left\{-\hspace{+0.2mm}\frac12\hspace{+0.25mm}\Big[c\hspace{+0.1mm}r^2+c'(r')^2\Big]\hspace{0.0mm}\right\}\hspace{-0.1mm},
\end{align}
where $\mathcal{I}_m(x)$ is the modified Bessel function of the first kind, and where the scalars $\{c_1,\,c,\,c'\!\!\hspace{0mm},\,a\}$ have been defined in Eqs.~\eqref{eq:c1}\hspace{+0.2mm}--\hspace{+0.2mm}\eqref{eq:a}.

The ground-state (`\hspace{+0.25mm}$\mathrm{GS}$\hspace{+0.40mm}') matrix element of the projected one-body density matrix can now be written as
\begin{align}
\label{eq:ground_state}
[\hspace{+0.2mm}\rho_\up^m(r,\,r')]_{\hspace{+0.1mm}\mathrm{GS}}\equiv\frac{\langle\hspace{+0.1mm}\Psi^{(\mathrm{GS})}\hspace{+0.2mm}|\hspace{+0.4mm}\rho_\up^{m}(r,\,r')\hspace{+0.4mm}|\hspace{+0.2mm}\Psi^{(\mathrm{GS})}\hspace{+0.1mm}\rangle}{\langle\hspace{+0.1mm}\Psi^{(\mathrm{GS})}\hspace{+0.3mm}|\hspace{+0.3mm}\Psi^{(\mathrm{GS})}\hspace{+0.1mm}\rangle}=\frac{\sum_{\hspace{+0.25mm}i,\,j}c_i^*\hspace{+0.30mm}[\hspace{+0.2mm}\rho_\up^m(r,\,r')]_{\mathbb{A}_i\mathbb{A}_j}c_j}{\sum_{\hspace{+0.25mm}i,\,j}c_i^*\hspace{+0.2mm}\mathbb{O}_{\mathbb{A}_i\mathbb{A}_j}c_j}\;.
\end{align}
Above, the second expression is obtained from the first by inserting two complete sets of explicitly correlated Gaussian basis states into both the numerator and denominator.  The $i^{t\hspace{-0.2mm}h}$ (real) coefficient of the total ground-state wave function in this basis is $c_i\equiv\langle\phi_{\mathbb{A}_i}|\hspace{+0.2mm}\Psi^{(\mathrm{GS})}\hspace{+0.1mm}\rangle$, and the overlap matrix element is~\cite{Suzuki_&_Varga_1998}
\begin{align}
\label{eq:overlap}
\mathbb{O}_{\mathbb{A}_i\hspace{+0.1mm}\mathbb{A}_j}\equiv\langle\phi_{\mathbb{A}_i}|\phi_{\mathbb{A}_j}\rangle=\frac{(2\pi)^{\hspace{+0.1mm}N}}{\mathrm{det\hspace{+0.2mm}[\mathbb{A}_\mathit{i}+\mathbb{A}_\mathit{j}\hspace{+0.1mm}]}}\;.
\end{align}
The indices $i$ and $j$ both run over the minimum number of (previously found) optimised basis states required to converge the ground-state energy at a given two-body binding energy $\varepsilon_b$.  While the equations in this and later subsections are written in terms of unsymmetrised basis states for clarity, these must be antisymmetrised to account for particle exchange (refer to Appendix D of Ref.~\cite{Laird_2024} for further details).

\begin{figure}[H]
\vspace{-3mm}
\hspace{-2.1mm}
\includegraphics[trim = 8mm 0mm 0mm 0mm, clip, scale = 0.68]{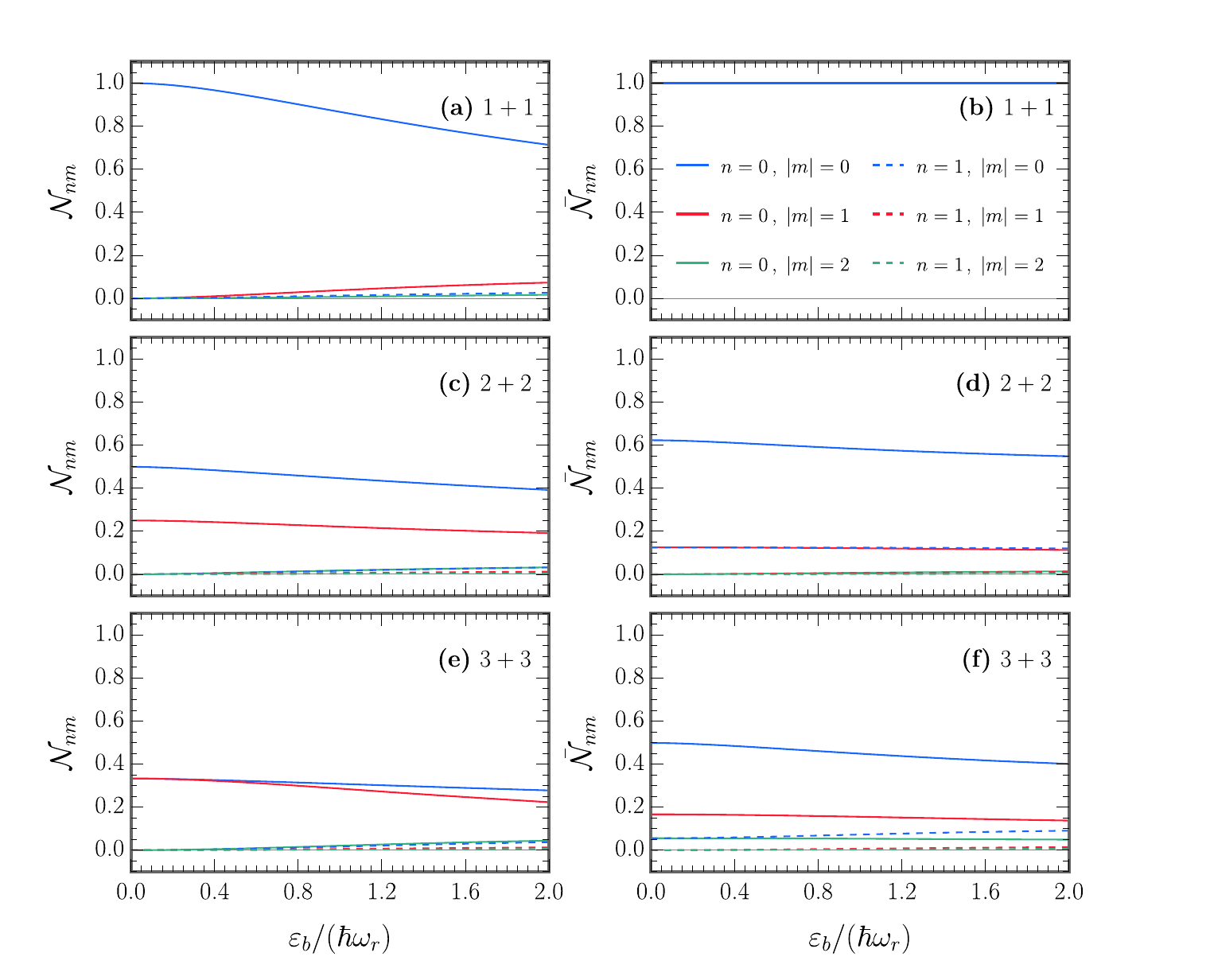}
\caption{Left panels: Ground-state occupation numbers (eigenvalues) of the one-body density matrix $\mathcal{N}_{nm}$~\eqref{eq:1BDM_decomposition} for (a) $1+1$, (c) $2+2$\hspace{+0.35mm}, and (e) $3+3$ fermions.  Right panels: Ground-state occupation numbers of the reduced two-body density matrix $\bar{\mathcal{N}}_{nm}$~\eqref{eq:2BDM_decomposition} for (b) $1+1$, (d) $2+2$\hspace{+0.35mm}, and (f) $3+3$ fermions.  The results are plotted as a function of the two-body binding energy $\varepsilon_b$ for (very nearly) zero effective range, $r_\mathrm{2D}/l_r^2=-\hspace{+0.3mm}0.001\approx0$.  Note in panel (b) that for any binding energy $\bar{\mathcal{N}}_{0\hspace{+0.1mm},\,0}=1$, while all other occupation numbers vanish.}
\label{fig:Fig_Occ-Nums}
\end{figure}

\vspace{+5.5mm}

At this point, the occupation numbers can be found by discretising the variables $r$ and $r'$ into grids of width $\Delta{r}\hspace{-0.5mm}$ and then finding the eigenvalues of $\sqrt{r}\,[\hspace{+0.2mm}\rho_\up^m(r,\,r')]_{\hspace{+0.1mm}\mathrm{GS}}\sqrt{r'}\Delta{r}$ for a given partial wave $m$.  The first such eigenvalue is $\mathcal{N}_{n\,=\,0\hspace{+0.1mm},\,m}\hspace{+0.2mm}$, the second is $\mathcal{N}_{n\,=\,1,\,m}\hspace{+0.2mm}$, and so on.  These results are shown in panels (a), (c), and (e) of Fig.~\ref{fig:Fig_Occ-Nums}.  In the non-interacting limit of $\varepsilon_b=0$, where the natural orbitals are the single-particle harmonic oscillator levels, they are straightforward to understand.  Due to the antisymmetry of the wave function same-spin ferm-\linebreak ions must occupy different single-particle levels.  For $1+1$ fermions the spin-up atom is in the $n=m=0$ ground state, which has an occupation number of $\mathcal{N}_{0\hspace{+0.1mm},\,0}=1$ due to the normalisation condition \eqref{eq:1BDM_occ-num_norm-cond}, while all other occupation numbers are zero.  For $2+2$ fermions the second spin-up atom is equally distributed between the two degenerate first excited states with $n=0$ and $m=\pm1$ --- leading to three finite occupation numbers, $\mathcal{N}_{0\hspace{+0.1mm},\,0}=1/2$ and $\mathcal{N}_{0\hspace{+0.1mm},\,\pm1}=1/4$.  In the $3+3$ case, the three lowest energy states contain one spin-up fermion each and thus the corresponding occupation numbers become $\mathcal{N}_{0\hspace{+0.1mm},\,0}=\mathcal{N}_{0\hspace{+0.1mm},\,\pm1}=1/3$, whereas all others vanish.

When the binding energy increases ($\varepsilon_b>0$) these finite occupation numbers decrease, while the occupation numbers of higher excited natural orbitals increase as one would generally expect.  However, for the range of interaction strengths covered by the energy spectra in Subsection~\ref{sec:Energy_Spectra} ($0\leq\varepsilon_b\leq2\hspace{+0.3mm}\hbar\omega_r$) this variation is not strong --- and the one-body density matrix can always be decomposed with a good level of accuracy by only including up to six natur-\linebreak al orbitals.  Such an observation suggests that we are never close to the deep Bose--Einstein condensation regime.  If we instead had a tight composite bosonic wave function, then its expansion into effective single-particle orbitals (the natural orbitals of $\rho_\up$) would require many terms~\cite{Blume_2011}\hspace{+0.1mm}.  In that case, many more occupation numbers would take on (small but) non-van-\linebreak ishing values, forcing a more significant reduction in the values of $\mathcal{N}_{0\hspace{+0.1mm},\,0}$ and $\mathcal{N}_{0\hspace{+0.1mm},\,\pm1}$ than what can be seen in Fig.~\ref{fig:Fig_Occ-Nums}.

\subsubsection{Two-Body Density Matrix}

The two-body density matrix in the first-quantised position representation is given by
\begin{align}
\label{eq:rho}
&\rho(\r_1,\,\r_1';\,\r_2,\,\r_2')=\left[\hspace{+0.20mm}\int\!\cdots\!\int{d}\r_1^\up{d}\r_2^\down\cdots{d}\r_{N-1}^\up{d}\r_N^\down\left|\Psi(\r_1^\up,\,\r_2^\down,\,\cdots,\,\r_{N-1}^\up,\,\r_N^\down)\right|^2\hspace{+0.10mm}\right]^{\hspace{+0.25mm}-1}\hspace{-2.9mm}\times\nn\\
&\hspace{-2mm}\int\!\cdots\!\int{d}\r_3^\up\hspace{+0.20mm}{d}\r_4^\down\cdots{d}\r_{N-1}^\up{d}\r_N^\down\Psi(\r_1,\,\r_2,\,\r_3^\up,\,\r_4^\down,\,\cdots,\,\r_{N-1}^\up,\,\r_N^\down)\hspace{+0.2mm}\Psi^*(\r_1',\,\r_2',\,\r_3^\up,\,\r_4^\down,\,\cdots,\,\r_{N-1}^\up,\,\r_N^\down)\;,
\end{align}
where the density $\Psi\Psi^*$ is integrated over all co-ordinates except those of one spin-$\up$ particle and one spin-$\down$ particle.  In two dimensions $\rho(\r_1,\,\r_1';\,\r_2,\,\r_2')$ is an eight-dimensional array, so we again need to reduce the number of degrees of freedom prior to diagonalisation.  To this end we follow Ref.~\cite{Blume_2011}\hspace{+0.1mm}, which considered the three-dimensional version of this problem, and transform from the co-ordinates of the individual atoms to the centre-of-mass and relative co-ordinates of the two spin-$\up$-spin-$\down$ pairs:  $\R=(\r_1+\r_2)/2$ and $\r=\r_1-\r_2\hspace{+0.2mm}$ (and their primed equivalents).  By setting $\r=\r'$\hspace{-0.5mm} we can then define the \textit{reduced} two-body density matrix as
\begin{align}
\label{eq:rho-red}
\rho_\mathrm{red}\hspace{+0.1mm}(\R,\,\R')=\int\!{d}\r\,\rho\!\left(\R+\frac{\r}{2},\,\R'+\frac{\r}{2};\,\R-\frac{\r}{2},\,\R'-\frac{\r}{2}\right)\hspace{+0.3mm},
\end{align}
which measures non-local correlations between pairs described by the \textit{same} relative-distance vector.

In analogy to the one-body density matrix, the reduced two-body density matrix can be expanded in terms of natural orbitals and occupation numbers:
\begin{align}
\label{eq:2BDM_decomposition}
\rho_\mathrm{red}\hspace{+0.1mm}(\R,\,\R')=\sum_{nm}\bar{\mathcal{N}}_{nm}\,\hspace{0mm}\bar{\chi}_{nm}^*(\R)\,\bar{\chi}_{nm}(\R')\;,
\end{align}
which have the normalisations,
\begin{subequations}
\begin{align}
\int\!d\R\,\bar{\chi}_{nm}^*(\R)\,\bar{\chi}_{n'\hspace{-0.15mm}m'}(\R)&=\delta_{nn'}\hspace{+0.2mm}\delta_{mm'}\;,\\
%\label{eq:1BDM_occ-num_norm-cond}
\sum_{nm}\bar{\mathcal{N}}_{nm}&=1\,\hspace{0mm}.
\end{align}
\end{subequations}
We again perform partial-wave projections according to Eq.~\eqref{eq:partial-wave_projections}: $\rho_\mathrm{red}\hspace{+0.1mm}(\R,\,\R')\to\rho_\mathrm{red}^m(R,\,R')$ with $R^{\hspace{+0.2mm}(}$$'$$^{)}\hspace{-0.1mm}\equiv|\R^{\hspace{+0.1mm}(}$$'$$^{)}|$.  The derivation of the ground-state matrix element of the \textit{projected reduced} two-body density matrix $[\hspace{+0.2mm}\rho_\mathrm{red}^m(R,\,R')]_{\hspace{+0.1mm}\mathrm{GS}}$ then follows identically to Eqs.~\eqref{eq:projected_1BDM}\hspace{+0.2mm}--\hspace{+0.2mm}\eqref{eq:ground_state} with only one minor change:  The vector of single-particle co-ordinates $\y$ must be replaced by $\y'\hspace{-0.5mm}$,
\begin{align}
\label{eq:new_y_vector}
\y=(\r_1^\up,\,\r_2^\down,\,\r_3^\up,\,\dots,\,\r_N^\down)\;\to\;\y'=(\R,\,\r,\,\r_3^\up,\,\dots,\,\r_N^\down)\;,
\end{align}
and therefore the transformation matrix $\mathbb{U}$ should be redefined appropriately,\;$\x=\mathbb{U}'\hspace{+0.1mm}\y'$~\cite{Blume_2011}.  The replacement matrix $\mathbb{U}'$ which takes the place of $\mathbb{U}$ is shown below for the various total\linebreak atom numbers (\hspace{+0.1mm}$N_\up\hspace{-0.2mm}+\hspace{+0.2mm}N_\down$\hspace{+0.1mm}) considered in this work (where the original $\mathbb{U}$ matrices were defin-\linebreak ed in Eq.~(A.2) of Ref.~\cite{Laird_2024}\hspace{+0.2mm}):
\begin{subequations}
\label{eq:new_U_matrices}
\begin{align}
&1+1:\quad\mathbb{U}
=
\left(\begin{array}{rrrrrr}
1 & -1 \\
\frac12 & \frac12 \\
\end{array}\right)
\;\to\;
\mathbb{U}'
=
\left(\begin{array}{rrrrrr}
0 & 1 \\
1 & 0 \\
\end{array}\right)\hspace{+0.4mm},\\
% \clearpage
%
&2+2:\quad\mathbb{U}
=
\left(\begin{array}{rrrrrr}
1 & -1 & 0 & 0\\
0 & 0 & 1 & -1  \\
\frac12 & \frac12 & -\frac12 & -\frac12  \\[3.5pt]
\frac14 & \frac14 & \frac14 & \frac14 \\
\end{array}\right)
\;\to\;
\mathbb{U}'
=
\left(\begin{array}{rrrrrr}
0 & 1 & 0 & 0  \\
0 & 0 & 1 & -1  \\
1 & 0 & -\frac12 & -\frac12 \\[3.5pt]
\frac12 & 0 & \frac14 & \frac14  \\
\end{array}\right)\hspace{+0.4mm},\\
&3+3:\quad\mathbb{U}
=
\left(\begin{array}{rrrrrr}
1 & -1 & 0 & 0 & 0 & 0 \\
0 & 0 & 1 & -1 & 0 & 0 \\
0 & 0 & 0 & 0 & 1 & -1 \\
\frac12 & \frac12 & -\frac12 & -\frac12 & 0 & 0 \\[3.5pt]
\frac14 & \frac14 & \frac14 & \frac14 & -\frac12 & -\frac12 \\[3.5pt]
\frac16 & \frac16 & \frac16 & \frac16 & \frac16 & \frac16 \\
\end{array}\right)
\;\to\;
\mathbb{U}'
=
\left(\begin{array}{rrrrrr}
0 & 1 & 0 & 0 & 0 & 0 \\
0 & 0 & 1 & -1 & 0 & 0 \\
0 & 0 & 0 & 0 & 1 & -1 \\
1 & 0 & -\frac12 & -\frac12 & 0 & 0 \\[3.5pt]
\frac12 & 0 & \frac14 & \frac14 & -\frac12 & -\frac12 \\[3.5pt]
\frac13 & 0 & \frac16 & \frac16 & \frac16 & \frac16 \\
\end{array}\right)\hspace{+0.25mm}.
\end{align}
\end{subequations}

The occupation numbers $\bar{\mathcal{N}}_{nm}$ are obtained as the eigenvalues of $\sqrt{R}\,[\hspace{+0.2mm}\rho_\mathrm{red}^m(R,\,R')]_{\hspace{+0.1mm}\mathrm{GS}}\sqrt{R'}\linebreak\times\Delta{R}$ and are displayed in panels (b), (d), and (f) of Fig.~\ref{fig:Fig_Occ-Nums}.  Although the values in the non-interacting limit ($\varepsilon_b=0$) are less intuitive than in the one-body case, they may be verified by comparing against analytically derived results.  In Appendix~\ref{sec:Analytical_Results_in_the_Non-Interacting_Limit} we detail these steps for the $2+2$ system as an example.  For increasing binding energy ($\varepsilon_b>0$) the occupation numbers from the reduced two-body density matrix follow the same qualitative trends as those from the one-body density matrix.  It may initially seem counter-intuitive that the largest eigenvalue $\bar{\mathcal{N}}_{0\hspace{+0.1mm},\,0}$ has a higher value in the absence of pairs ($\varepsilon_b=0$) than in the presence of pairs ($\varepsilon_b\gg0$).  However, this is directly due to the procedure used to eliminate degrees of freedom and define the quantity $\rho_\mathrm{red}\hspace{+0.1mm}(\R,\,\R')$ --- and was similarly observed in the three-dimensional case~\cite{Blume_2011}.

\subsubsection{Molecular Condensate Fraction}

For a trapped one-component Bose gas the condensate fraction becomes appreciable when the lowest eigenvalue of the one-body density matrix becomes of order unity.  For a two-com-\linebreak ponent Fermi gas, by contrast, none of the natural orbitals of the one-body density matrix can become macroscopically occupied due to the antisymmetry of the wave function under particle exchange.  A significant condensate fraction can only arise when bosonic pairs are formed, and hence, such insight must instead come from an analysis of the two-body density matrix.

Due to the elimination of degrees of freedom as explained above, the value of $\bar{\mathcal{N}}_{0\hspace{+0.1mm},\,0}$ by it-\linebreak self cannot provide a direct measure of the fraction of condensed pairs.  Rather, the system is said to be condensed when the lowest natural orbital of the reduced two-body density matrix becomes macroscopically occupied --- that is, when $\bar{\mathcal{N}}_{0\hspace{+0.1mm},\,0}$ greatly exceeds all other $\bar{\mathcal{N}}_{nm}$~\cite{Blume_2011}.  Correspondingly, we define the condensate fraction $\mathcal{N}_\mathrm{cond}$ in two dimensions as follows:
\begin{align}
\label{eq:Ncond}
\mathcal{N}_\mathrm{cond}=1-\frac{\mathrm{max}\left(\hspace{+0.2mm}\sum_{\,m\,=\,\pm\,\dots}\bar{\mathcal{N}}_{nm}\hspace{+0.2mm}\right)}{\bar{\mathcal{N}}_{0\hspace{+0.1mm},\,0}}\,\,[(n,\,m)\neq(0,\,0)]\;,
\end{align}
which is analogous to the three-dimensional expression appearing in Eq.~(16) of Ref.~\cite{Blume_2011}.  Here, the sum applies when $m>0$ since the occupation numbers are degenerate for given $|m|$.  In the BEC limit ($\varepsilon_b\gg0$) the second term on the right-hand-side of Eq.~\eqref{eq:Ncond} above becomes small (because the numerator becomes \textit{very} small) and $\mathcal{N}_\mathrm{cond}$ approaches unity.  (Note that although $\bar{\mathcal{N}}_{0\hspace{+0.1mm},\,0}$ decreases as the two-body binding energy increases, in the BEC limit it is the only sizeable occupation number and a relatively large number of other $\bar{\mathcal{N}}_{nm}$ take on non-vanishing but very small values~\cite{Blume_2011}.)  In the non-interacting limit ($\varepsilon_b=0$) the second term in Eq.~\eqref{eq:Ncond} becomes of order unity and $\mathcal{N}_\mathrm{cond}$ approaches zero for large numbers of particles, while for small particle numbers $\mathcal{N}_\mathrm{cond}$ becomes a fraction less than one.

\begin{figure}[ht]
\begin{center}
\hspace{0mm}
\includegraphics[trim = 7.5mm 0mm 0mm 0mm, clip, scale = 0.68]{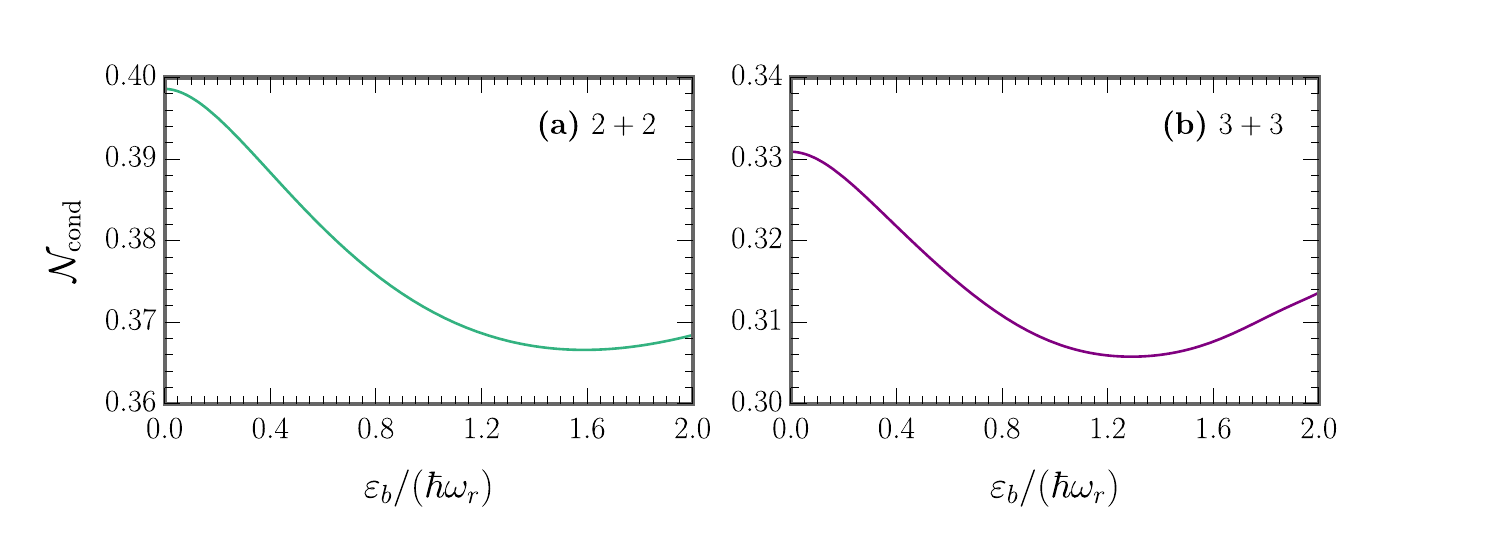}
\caption{The condensate fraction $\mathcal{N}_\mathrm{cond}$~\eqref{eq:Ncond} as a function of the two-body binding energy $\varepsilon_b$ for (a) $2+2$ and (b) $3+3$ fermions in the ground state.  In both cases the effective range is very close to zero, $r_\mathrm{2D}/l_r^2=-\hspace{+0.2mm}0.001\approx0$.}
\label{fig:Fig_Ncond}
\end{center}
\end{figure}

In Fig.~\ref{fig:Fig_Ncond} we plot the condensate fractions for the $2+2$ and $3+3$ Fermi systems,
\begin{subequations}
\begin{align}
\label{eq:Ncond_2+2_&_3+3}
&2+2:\quad\mathcal{N}_\mathrm{cond}=1-\frac{\bar{\mathcal{N}}_{1,\,0}+\bar{\mathcal{N}}_{0\hspace{+0.1mm},\,+1}+\bar{\mathcal{N}}_{0\hspace{+0.1mm},\,-1}}{\bar{\mathcal{N}}_{0\hspace{+0.1mm},\,0}}\hspace{+0.2mm}\,,\\
&3+3:\quad\mathcal{N}_\mathrm{cond}=1-\frac{\bar{\mathcal{N}}_{0\hspace{+0.1mm},\,+1}+\bar{\mathcal{N}}_{0\hspace{+0.1mm},\,-1}}{\bar{\mathcal{N}}_{0\hspace{+0.1mm},\,0}}\hspace{+0.2mm}\,,
\end{align}
\end{subequations}
as functions of the interaction strength $\varepsilon_b$.  (Recall that for $2+2$ fermions $\bar{\mathcal{N}}_{1,\,0}$ is degenerate with $\bar{\mathcal{N}}_{0\hspace{+0.1mm},\,\pm1}$ in Fig.~\ref{fig:Fig_Occ-Nums}.)  The results for both atom numbers are qualitatively and quantitatively similar.  For binding energies of  $\varepsilon_b\gtrsim2\hspace{+0.3mm}\hbar\omega_r$\hspace{+0.1mm}, \hspace{+0.2mm}$\mathcal{N}_\mathrm{cond}$ appears to be increasing in both cases and should become maximal (equal to one) when the fermions are tightly bound into diatomic molecules.  This reinforces our earlier conclusion --- made when discussing the values of $\mathcal{N}_{nm}$ in Fig.~\ref{fig:Fig_Occ-Nums} --- that we are never close to the deep BEC regime for our considered range of interaction strengths.  For binding energies of  $\varepsilon_b\lesssim2\hspace{+0.3mm}\hbar\omega_r$\hspace{+0.1mm}, the non-monotonic dependence of $\mathcal{N}_\mathrm{cond}$ stands in contrast to that observed in three-dimensional trapped~\cite{Blume_2011} and homogeneous~\cite{Astrakharchik_2005} systems where the condensate fraction always increases monotonically with interaction strength.  However, it is important to note that the variation of $\mathcal{N}_\mathrm{cond}$ within the non-monotonic regime is relatively small in both panels of Fig.~\ref{fig:Fig_Ncond}.  Given that $\mathcal{N}_\mathrm{cond}$ was first defined  in Ref.~\cite{Blume_2011} to describe trapped few-fermion systems (with $N\leq6$) in three dimensions, extending that definition to two dimensions here is certainly of interest.  Nonetheless, our results suggest that it has limited interpretability when applied to such small numbers of atoms.

\subsection{Momentum Distributions}
\label{sec:Momentum_Distributions}

The momentum distribution of the spin-$\up$ atoms is given by the Fourier transform of the one-body density matrix defined in Eq.~\eqref{eq:rho-up}:
\begin{align}
\label{eq:1B-FT}
n_\up(\k)=\frac{1}{(2\pi)^2}\int\!\int{d}\r\,{d}\r'\hspace{-0.2mm}\rho_\up(\r,\,\r')\,\hspace{+0.3mm}\mathrm{exp}\hspace{-0.1mm}\left[-\hspace{+0.2mm}i\hspace{+0.1mm}\k^T\hspace{-0.2mm}(\r\,\hspace{+0.1mm}-\,\r\hspace{+0.2mm}')\hspace{-0.1mm}\right]\hspace{-0.1mm}\,.
\end{align}
It is straightforward to prove that Eq.~\eqref{eq:1B-FT} is equivalent to
\begin{align}
n_\up(\k)=\sum_{nm}\mathcal{N}_{nm}\,|\hspace{+0.1mm}\widetilde\chi_{nm}(\k)\hspace{+0.2mm}|^{\hspace{+0.1mm}2}\hspace{-0.5mm},
\end{align}
where
\begin{align}
\widetilde\chi_{nm}(\k)=\frac{1}{2\pi}\int{d}\r\,\chi_{nm}(\r)\;\mathrm{exp}\hspace{-0.1mm}\left(-\hspace{+0.2mm}i\hspace{+0.1mm}\k^T\hspace{-0.3mm}\r\hspace{+0.1mm}\right)
\end{align}
is the Fourier transform of the natural orbitals introduced in Eq.~\eqref{eq:1BDM_decomposition}.  In order to obtain an analytical expression for the matrix elements of Eq.~\eqref{eq:1B-FT} in the explicitly correlated Gaussian basis, we can use the result for $[\hspace{+0.2mm}\rho_\up(\r,\,\r')]_{\mathbb{A}\hspace{+0.1mm}\mathbb{A}'}$ shown in Eq.~\eqref{eq:final_rho-up}:
\begin{align}
\label{eq:1B-FT_sub}
[n_\up(\k)]_{\mathbb{A}\mathbb{A}'}=\frac{c_1}{(2\pi)^2}\int\!\int{d}\r\,{d}\r'\hspace{+0.2mm}\mathrm{exp}\hspace{+0.1mm}\!\left\{-\hspace{+0.2mm}\frac12\hspace{+0.25mm}\Big[c\hspace{+0.1mm}\r^{\hspace{+0.3mm}2}+c'(\r'\hspace{+0.1mm})^{\hspace{+0.1mm}2}-a\hspace{+0.1mm}\r^T\hspace{-0.3mm}\r'\Big]\hspace{0.0mm}\right\}\mathrm{exp}\hspace{-0.1mm}\left[i\hspace{+0.1mm}\k^T\hspace{-0.2mm}(\r\hspace{+0.2mm}'\hspace{-0.3mm}-\,\r)\hspace{-0.1mm}\right]\hspace{-0.1mm}\,.
\end{align}
By defining $\mathbf{X}=\r'\hspace{-0.5mm}-\r\hspace{+0.1mm}$ the equation above becomes
\begin{align}
\label{eq:1B-FT_sub_again}
[n_\up(\k)]_{\mathbb{A}\mathbb{A}'}=\frac{c_1}{(2\pi)^2}\int\!\int{d}\r\,{d}\X\;\hspace{+0.1mm}\mathrm{exp}\hspace{-0.1mm}\left[\frac12\hspace{+0.25mm}\Big(g_1\r^{\hspace{+0.3mm}2}+g_2\hspace{+0.2mm}\X^{\hspace{+0.1mm}2}+g_3\hspace{+0.2mm}\r^T\hspace{-0.1mm}\X\Big)\hspace{-0.1mm}\right]\hspace{+0.2mm}\mathrm{exp}\hspace{-0.1mm}\left(i\hspace{+0.1mm}\k^T\hspace{-0.1mm}\X\hspace{+0.2mm}\right)\hspace{-0.1mm}\,,
\end{align}
which depends on
\begin{subequations}
\begin{align}
&g_1=a-c-c\hspace{+0.2mm}'\,\hspace{-0.8mm},\\
&g_2=-\hspace{+0.2mm}c\hspace{+0.2mm}'\,\hspace{-0.8mm},\\
&g_3=a-2c\hspace{+0.2mm}'\,\hspace{-0.8mm}.
\end{align}
\end{subequations}
The integral over $\r$ can be performed analytically for $g_1\hspace{-0.4mm}<0$:
\begin{align}
\label{eq:1B-FT_sub_r_int}
[n_\up(\k)]_{\mathbb{A}\mathbb{A}'}=-\hspace{+0.2mm}\frac{c_1}{2\pi\hspace{+0.1mm}{g}_1}\int{d}\X\;\hspace{+0.1mm}\mathrm{exp}\hspace{-0.1mm}\left(\frac{1}{2}g_4\hspace{+0.3mm}\X^{\hspace{+0.1mm}2}\hspace{-0.1mm}\right)\hspace{+0.25mm}\mathrm{exp}\hspace{-0.1mm}\left(i\hspace{+0.1mm}\k^T\hspace{-0.1mm}\X\hspace{+0.2mm}\right)\hspace{-0.1mm}\,,
\end{align}
which contains
\begin{align}
g_4=g_2-g_3^2\hspace{+0.2mm}/\hspace{+0.2mm}(4\hspace{+0.1mm}g_1)\;.
\end{align}
Subsequently, the integral over $\X$ can be analytically carried out for $g_4\hspace{-0.15mm}<0$:
\begin{align}
\label{eq:1B-FT_sub_X_int}
[n_\up(\k)]_{\mathbb{A}\mathbb{A}'}\equiv[n_\up(k)]_{\mathbb{A}\mathbb{A}'}=\frac{c_1}{{g}_1g_4}\,\mathrm{exp}\hspace{-0.1mm}\left(\frac{1}{2\hspace{+0.1mm}g_4}\hspace{+0.3mm}k^{\hspace{+0.1mm}2}\hspace{-0.1mm}\right)\hspace{-0.1mm}\,,\quad{k}\equiv|\hspace{+0.1mm}\k\hspace{+0.3mm}|\;,
\end{align}
where the coefficient $c_1/({g}_1g_4)$ can be both positive and negative.

Analogously, the momentum distribution corresponding to the centre-of-mass motion of spin-$\up$-spin-$\down$ pairs is given by the Fourier transform of the reduced two-body density matrix defined in Eq.~\eqref{eq:rho-red}:
\begin{align}
\label{eq:2B-FT}
n\hspace{+0.1mm}(\K)=\frac{1}{(2\pi)^2}\int\!\int{d}\R\,\hspace{+0.2mm}{d}\R'\rho_\mathrm{red}(\R,\,\R')\,\hspace{+0.3mm}\mathrm{exp}\hspace{-0.1mm}\left[-\hspace{+0.2mm}i\hspace{+0.1mm}\K^{\hspace{+0.1mm}T}\hspace{-0.2mm}(\R\,\hspace{+0.1mm}-\,\R\hspace{+0.2mm}')\hspace{-0.1mm}\right]\hspace{-0.1mm}\,.
\end{align}

\begin{figure}[H]
\vspace{-3mm}
\hspace{-5.5mm}
\includegraphics[trim = 0mm 0mm 0mm 0mm, clip, scale = 0.679]{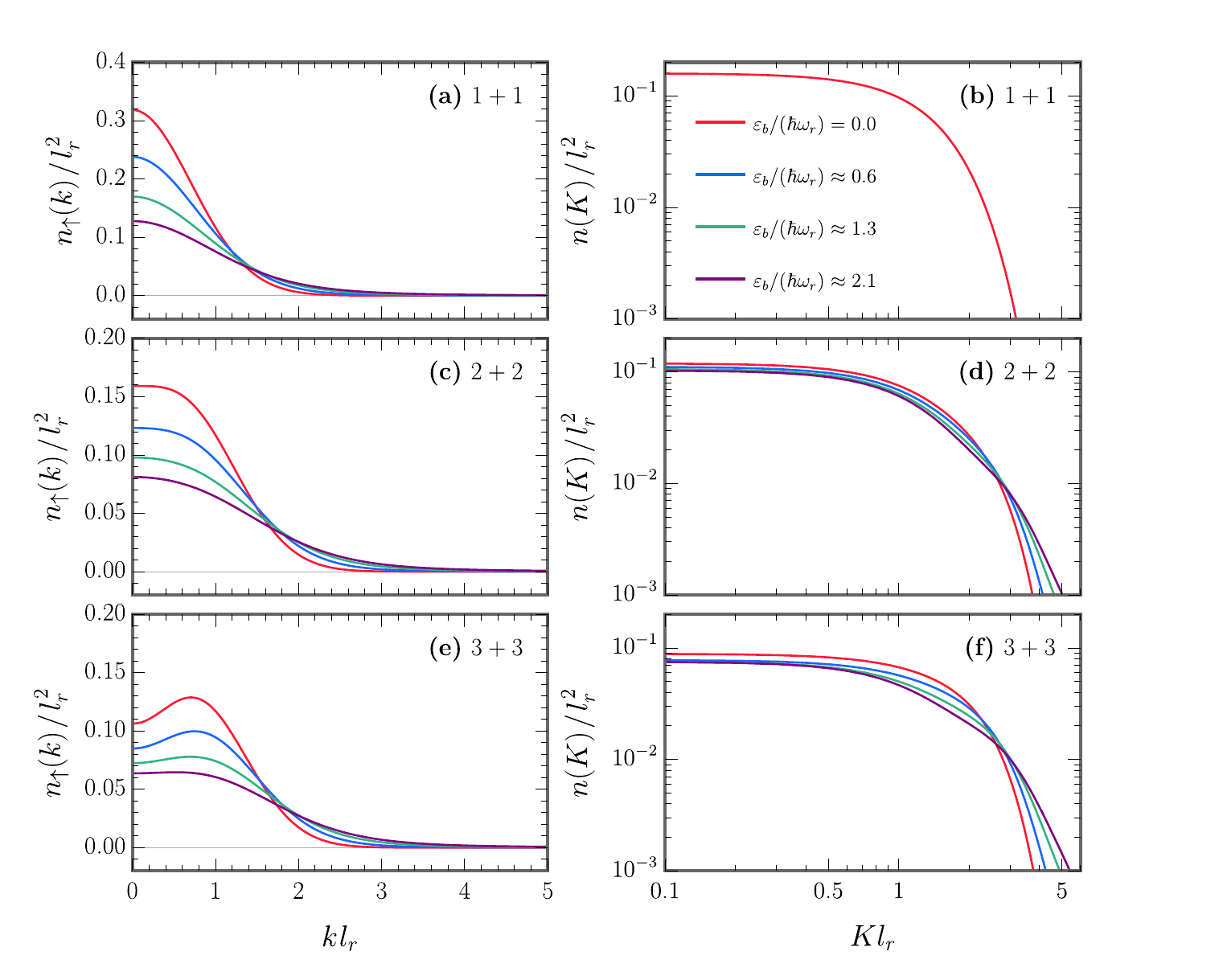}
\caption{Left panels:  The momentum distribution $n_\up(k)$~\eqref{eq:1B-FT} associated with the motion of spin-$\up$ atoms for (a) $1+1$, (c) $2+2$\hspace{+0.35mm}, and (e) $3+3$ fermions in the ground state.  Right panels:  The momentum distribution $n\hspace{+0.1mm}(K)$~\eqref{eq:2B-FT} associated with the cen-\linebreak tre-of-mass motion of spin-$\up$-spin-$\down$ pairs for (b) $1+1$, (d) $2+2$\hspace{+0.35mm}, and (f) $3+3$ fer-\linebreak mions in the ground state [note the log-log scale].  The differently coloured lines correspond to different binding energies $\varepsilon_b$\hspace{+0.1mm}, while the effective range is fixed for all lines to very nearly zero, $r_\mathrm{2D}/l_r^2=-\hspace{+0.2mm}0.001\approx0$.  By construction, the two-body results for $1+1$ fermions in panel (b) are the same at all values of $\varepsilon_b$.}
\label{fig:Fig_Mom_Dist}
\end{figure}

\vspace{5.5mm}

\noindent Here, we utilise the symbol $\K$ instead of $\k$ to distinguish the momentum vector associated with the position vector of a pair from that of an atom.  As with the calculation of the occupation numbers, the derivation of an analytical expression for $[n\hspace{+0.1mm}(K)]_{\mathbb{A}\mathbb{A}'}$ follows identically to the one above for $[n_\up(k)]_{\mathbb{A}\mathbb{A}'}$ with a single minor adjustment:  The transformation matrix $\mathbb{U}$ used to compute $\{c_1,\,c,\,c'\!\!\hspace{0mm},\,a\}$ in Eq.~\eqref{eq:1B-FT_sub} should be replaced by $\mathbb{U}'$\hspace{-0.5mm}, as explained in the text around Eqs.~\eqref{eq:new_y_vector}\hspace{+0.2mm}--\hspace{+0.2mm}\eqref{eq:new_U_matrices}.  Note that our treatment in this subsection was inspired by a complementary three-dimensional calculation in Ref.~\cite{Blume_2011} (see Appendix~A therein).

In Fig.~\ref{fig:Fig_Mom_Dist} we present the momentum distributions $[n_\up(k)]_{\hspace{+0.1mm}\mathrm{GS}}$ and $[n\hspace{+0.1mm}(K)]_{\hspace{+0.1mm}\mathrm{GS}}$ for the ground state which were calculated by replacing $[\hspace{+0.2mm}\rho_\up^m(r,\,r')]_{\mathbb{A}_i\mathbb{A}_j}$\hspace{-0.5mm} with $[n_\up(k)]_{\mathbb{A}_i\mathbb{A}_j}$\hspace{-0.5mm} and $[n\hspace{+0.1mm}(K)]_{\mathbb{A}_i\mathbb{A}_j}$\hspace{-0.5mm} in Eq.~\eqref{eq:ground_state}.  In the non-interacting thermodynamic limit the momentum distribution for a single spin component features a `\hspace{+0.1mm}step\hspace{+0.1mm}' at the Fermi momentum.  However, when there are only very few atoms this step becomes `\hspace{+0.1mm}smeared out\hspace{+0.2mm}' with a width determined by the radial harmonic trapping frequency ${k}_{\hspace{-0.2mm}r}\sim1/{l}_r=\sqrt{{m}\omega_r/\hbar}$\hspace{+0.2mm}, as shown in panels (a), (c), and (e).  Interestingly, $n_\up(k)$ adopts a distinct shape for each number of fermions, with the non-monotonicity in the $3+3$ case likely resulting from finite-size effects of the trap.  By contrast, the distribution $n\hspace{+0.1mm}(K)$\linebreak displayed in panels (b), (d), and (f) varies little with either particle number or binding energy.  For the particular case of $1+1$ fermions [Fig.~\ref{fig:Fig_Mom_Dist}(b)] $n\hspace{+0.1mm}(K)$ shows no dependence on the binding energy, mirroring the behaviour of the occupation numbers in Fig.~\ref{fig:Fig_Occ-Nums}(b).  In three dimensions~\cite{Blume_2011} $n\hspace{+0.1mm}(K)$ was found to exhibit two distinct features in the limit of small positive scattering length that could be associated with the condensation of pairs: a feature at smaller $K$ corresponding to the momentum distribution of non-interacting composite bosons of mass $\hspace{+0.2mm}2m$, and a feature at larger $K$ corresponding to the internal structure of the bosons.  For our largest considered binding energy $\varepsilon_b\approx2.1\hspace{+0.3mm}\hbar\omega_r$ we begin to see a `\hspace{+0.1mm}shoulder\hspace{+0.2mm}' emerging at larger $K$ that resembles this phenomenon, however, it is much less pronounced.  This suggests --- consistent with the previous subsections --- that we remain far from the deep BEC regime.

\subsection{Radial and Pair Distribution Functions}
\label{sec:Radial_and_Pair_Distribution_Functions}

As well as the density matrices, any \textit{local} structural property $P(r)$ of the $N\hspace{-0.3mm}$-body system can be calculated from the wave function as follows~\cite{von_Stecher_2008,Blume_2011,Bradly_2014}:
\begin{align}
\label{eq:gen_struct_prop}
P(r)=\int\!{d}\r'\hspace{+0.2mm}\frac{\delta(r-r')}{2\pi{r'}}\int\!{d}^{2N}\hspace{-0.3mm}\x\,\delta(\r'-\x)\hspace{-0.1mm}\,|\Psi(\x)\hspace{+0.1mm}|^{\hspace{+0.1mm}2}\hspace{-0.5mm}.
\end{align}
Here, $\r$ (and $\r'$\hspace{+0.2mm})  is a co-ordinate describing the property of interest and $\x$ is a generalised set\linebreak of co-ordinates, such as the set of $N$ Jacobi position vectors defined in Appendix~A of Ref.~\cite{Laird_2024} which includes the centre-of-mass position.  We can define the averaged radial one-body density $P_\up(r)$ by setting $\r=\r_1$ in Eq.~\eqref{eq:gen_struct_prop},\footnote{\hspace{+0.3mm}Because the Fermi gases of interest are spin-balanced, the radial one-body densities for the spin-up and spin-down atoms are equal, $P_\up(r)=P_\down(r)$.  In addition, since we consider only the sector of zero total orbital angular momentum, $P_\up(r)$ is radially (circularly) symmetric.} and also the averaged radial pair distribution function $P_{\up\down}(r)$ by setting $\r=\r_1-\r_2$\hspace{+0.1mm}.  These quantities are normalised such that
\begin{align}
\label{eq:norms}
2\pi\int_0^\infty\!\!\!\!d\hspace{-0.3mm}r\,\hspace{-0.5mm}r\,\hspace{-0.4mm}P_\up(r)=1\quad\mathrm{and}\quad2\pi\int_0^\infty\!\!\!\!d\hspace{-0.3mm}r\,\hspace{-0.5mm}r\,\hspace{-0.4mm}P_{\up\down}(r)=1\,.
\end{align}
The value of $2\pi{r}P_\up(r)\hspace{+0.5mm}d\hspace{-0.3mm}r$ therefore equals the probability of locating a particle at a distance between $r$ and $r+d\hspace{-0.3mm}r$ from the centre of the trap.  Likewise, the value of $2\pi{r}P_{\up\down}(r)\hspace{+0.5mm}d\hspace{-0.3mm}r$ equates to the probability of locating a spin-up particle and a spin-down particle at a distance between $r$ and $r+d\hspace{-0.3mm}r$ from each other.

We compute the ground-state matrix element $[P_\sigma(r)]_{\hspace{+0.1mm}\mathrm{GS}}$ ($\sigma\equiv\;\up$ or $\up\down$) in a similar manner to Eq.~\eqref{eq:ground_state}.  In the explicitly correlated Gaussian basis, the matrix elements for arbitrary one- and two-body operators are respectively given by
\begin{subequations}
\begin{align}
\label{eq:1-body}
\langle\phi_{\mathbb{A}_i}|\hspace{-0.2mm}\,V(\r_k)\,|\phi_{\mathbb{A}_j}\rangle&=\hspace{+0.2mm}\mathbb{O}_{\mathbb{A}_i\mathbb{A}_j}\,\frac{b_k}{2\pi}\int{d}\mathbf{r}\,V(\mathbf{r})\hspace{+0.3mm}\,\mathrm{exp}\hspace{-0.3mm}\left(-\hspace{+0.2mm}\frac12b_kr^2\right)\hspace{-0.2mm}\,,\\
\label{eq:2-body}
\langle\phi_{\mathbb{A}_i}|\hspace{-0.2mm}\,V(\r_k-\r_l)\,|\phi_{\mathbb{A}_j}\rangle&=\hspace{+0.2mm}\mathbb{O}_{\mathbb{A}_i\mathbb{A}_j}\,\frac{b_{kl}}{2\pi}\int{d}\mathbf{r}\,V(\mathbf{r})\hspace{+0.3mm}\,\mathrm{exp}\hspace{-0.3mm}\left(-\hspace{+0.2mm}\frac12b_{kl}r^2\right)\hspace{-0.2mm}\,,
\end{align}
\end{subequations}
where
\begin{subequations}
\begin{align}
&\frac{1}{b_k}=\big[\mathbf{\bm{\omega}}^{(k)}\big]^T\hspace{-0.5mm}(\mathbb{A}_i+\mathbb{A}_j)^{-1}\bm{\omega}^{(k)}\hspace{-0.2mm}\,,\quad\big[\mathbf{\bm{\omega}}^{(k)}\big]_p=\hspace{+0.2mm}(\mathbb{U}^{-1})_{kp}\;,\\
&\frac{1}{b_{kl}}=\big[\mathbf{\bm{\omega}}^{(kl)}\big]^T\hspace{-0.5mm}(\mathbb{A}_i+\mathbb{A}_j)^{-1}\bm{\omega}^{(kl)}\hspace{-0.2mm}\,,\quad\big[\mathbf{\bm{\omega}}^{(ij)}\big]_p=\hspace{+0.2mm}(\mathbb{U}^{-1})_{ip}-(\mathbb{U}^{-1})_{jp}\;,
\end{align}
\end{subequations}
and $p=1,\,\dots,\,N$\cite{Suzuki_&_Varga_1998,Laird_2024}\hspace{+0.1mm}.  Correspondingly, we substitute $V(\r_k)=\delta(\r-\r_k)$ into Eq.~\eqref{eq:1-body} to evaluate $[P_{\up}(r)]_{\mathbb{A}_i\mathbb{A}_j}$ and $V(\r_k-\r_l)=\delta(\r-\r_k-\r_l)$ into Eq.~\eqref{eq:2-body} to determine $[P_{\up\down}(r)]_{\mathbb{A}_i\mathbb{A}_j}$.

\begin{figure}[H]
\begin{center}
\vspace{-0.5mm}
\hspace{0mm}
\includegraphics[trim = 8mm 0mm 0mm 0mm, clip, scale = 0.68]{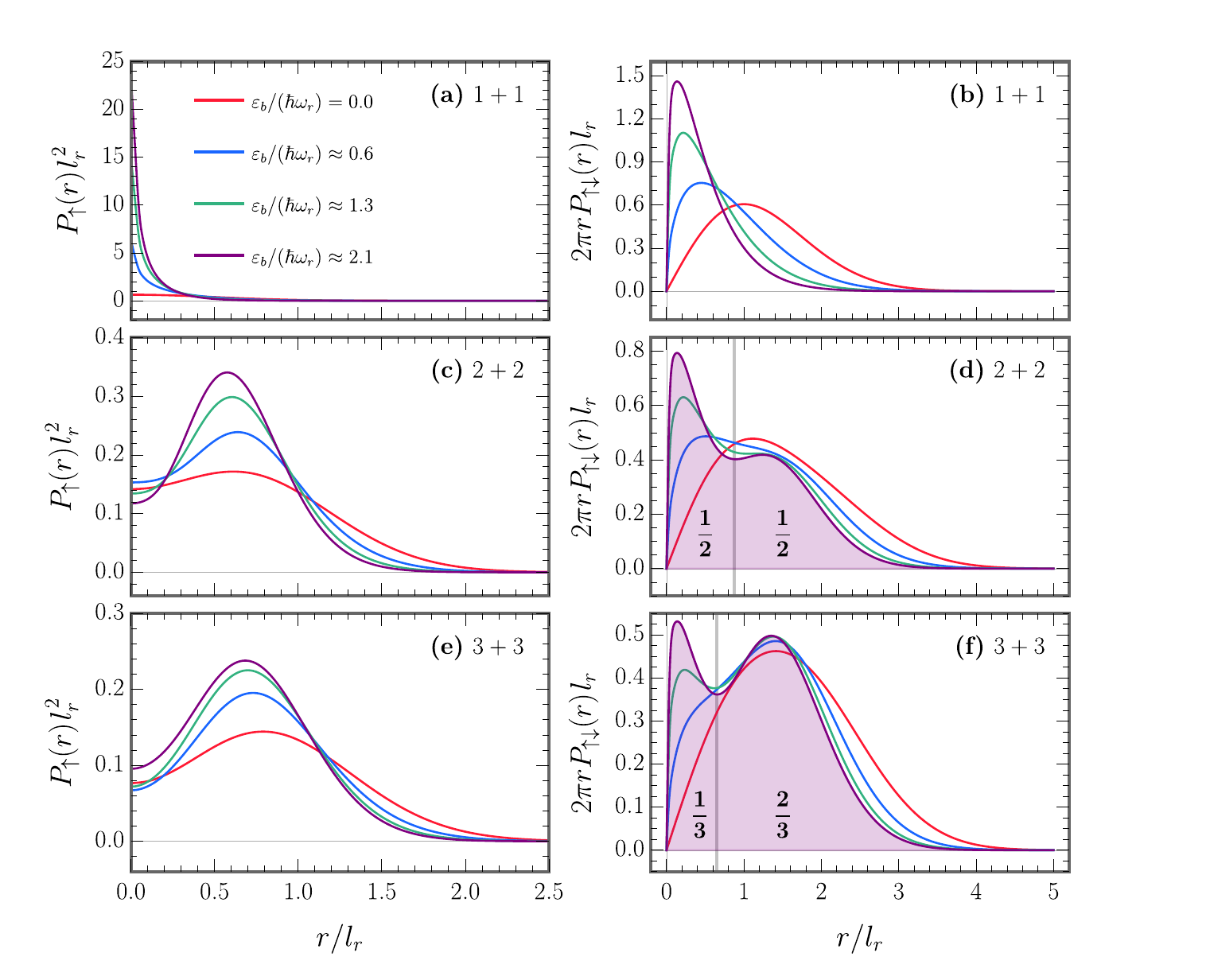}
\caption{Left panels:  The radial one-body density $P_\up(r)$ for (a) $1+1$, (c) $2+2$\hspace{+0.35mm}, and (e) $3+3$ fermions in the ground state.  Right panels:  The (scaled) radial pair distri-\linebreak bution function $P_{\up\down}(r)$ for (b) $1+1$, (d) $2+2$\hspace{+0.35mm}, and (f) $3+3$ fermions in the ground state.  The results are shown for a variety of binding energies ($\varepsilon_b$) at close to zero ef-\linebreak fective range ($r_\mathrm{2D}/l_r^2=-\hspace{+0.2mm}0.001\approx0$).  The bold fractions indicate the (approximate) shaded area under the curve on either side of the grey vertical line for $\varepsilon_b\approx2.1\hspace{+0.1mm}\hbar\omega_r$\hspace{+0.1mm}, as discussed in the main text.}
\label{fig:Fig_RD_&_UDPD_new}
\end{center}
\end{figure}

\vspace{+1.5mm}

Our results for the radial one-body density are shown in panels (a), (c), and (e) of Fig.~\ref{fig:Fig_RD_&_UDPD_new}.  When $N_\up=1$ the peak density is located at the centre of the trap (at $r=0$).  However, when $N_\up=2$ and $N_\up=3$ the peak density shifts to a finite value of $r$ on the order of the radial harmonic trap length $l_r$\hspace{+0.1mm}, which sets the average interparticle spacing.  This shift from zero to finite $r$ as the number of fermions increases is a signature of both the (residual) shell structure of the two-dimensional harmonic oscillator and the Pauli exclusion principle.  For $N_\up=1$ the first harmonic oscillator shell in the non-interacting ground state is fully occupied; whereas for $N_\up=2$ and $N_\up=3$ fermions occupy the first two shells in the non-interacting ground state, leading to similar results in both cases.  In order to accommodate both the radial symmetry and Pauli repulsion between identical spins in these larger systems, $P_\up(r)$ maintains a single peak that shifts outwards from the trap centre.

Panels (b), (d), and (f) of Fig.~\ref{fig:Fig_RD_&_UDPD_new} show our results for the (scaled) radial pair distribution function.  At binding energies of $\varepsilon_b\gtrsim\hspace{+0.2mm}\hbar\omega_r$ and when there is more than one particle per spin state, $rP_{\up\down}(r)$ develops a clear two-peak structure similar to what has been observed in three dimensions~\cite{von_Stecher_2008,Blume_2011}.  The peak at smaller $r$ (around $0.1l_r$) signifies the formation of weakly bound dimers, while the peak at larger $r$ (between $1l_r$ and $2\hspace{+0.3mm}l_r$) is set by the dimer-dimer distance which is longer due to Pauli repulsion between same-spin fermions.  The $2+2$ system\linebreak has two such small interspecies distances (the distance between a spin-up and spin-down particle within a pair) and two large interspecies distances (the distance between a spin-up and spin-down particle in different pairs).  Accordingly, if we integrate $P_{\up\down}(r)$ for $N_\up=2$ from zero up to the $r$ value where $rP_{\up\down}(r)$ features a minimum, then we find that the probability of forming a molecule (of being at short distances) is $\sim1/2$~\cite{von_Stecher_2008}.  On the other hand, the $3+3$ system has three small interspecies distances and six large interspecies distances --- and performing a similar integration confirms the probability of forming a molecule to be $\sim1/3$.  These probabilities are indicated in the figure.  If we were to access the deep BEC regime $\varepsilon_b\gg2\hspace{+0.3mm}\hbar\omega_r$\hspace{+0.1mm}, then the peak at smaller $r$ would become taller and narrower, while the peak at larger $r$ would become shorter and broader, with the pair density in between them reducing almost to zero --- and the fractions mentioned above would become exactly $1/2$ and $1/3$~\cite{von_Stecher_2008}.  The reason why the scaled pair distribution function vanishes for $r\to0$ is because we are using a finite-range interaction potential, such that unlike spins cannot approach each other at distances $\lesssim{r}_0$.  If we had instead considered zero-range interactions, then the amplitude of $rP_{\up\down}(r)$ would have been finite at $r=0$~\cite{Blume_2011,Yan_2015}.

\begin{figure}[H]
\begin{center}
\vspace{7mm}
\hspace{6mm}
\includegraphics[trim = 8mm 0mm 0mm 0mm, clip, scale = 0.68]{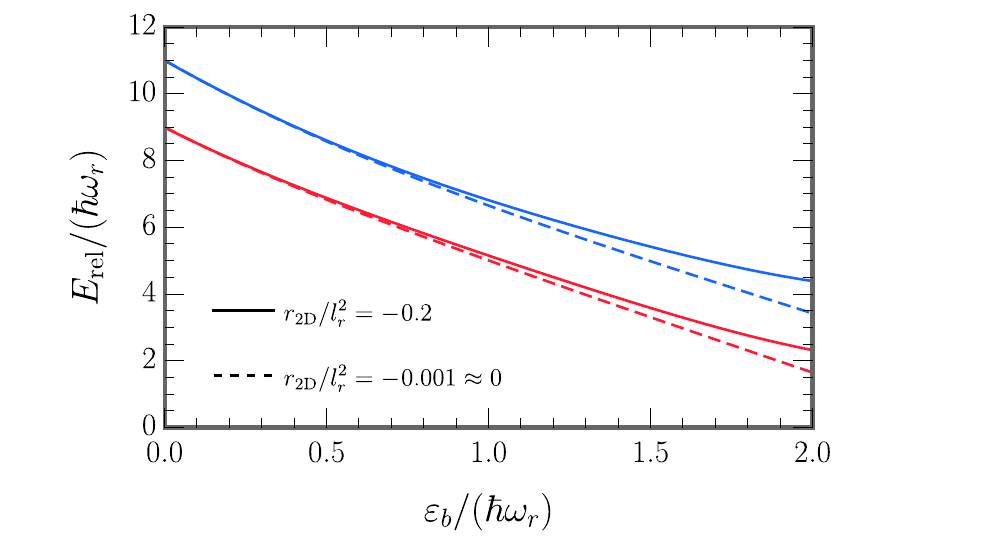}
\caption{The relative energy of the ground [red] and first excited state [blue] as a function of the two-body binding energy for $3+3$ fermions in the monopole sector of zero total orbital angular momentum.  Solid lines correspond to an effective range of $r_\mathrm{2D}/l_r^2=-\hspace{+0.2mm}0.2$ and dashed lines to $r_\mathrm{2D}/l_r^2=-\hspace{+0.2mm}0.001\approx0$.}
\label{fig:Fig_ES_2}
\end{center}
\end{figure}

\vspace{-6mm}

\begin{figure}[H]
\hspace{-2.1mm}
\includegraphics[trim = 8mm 0mm 0mm 0mm, clip, scale = 0.68]{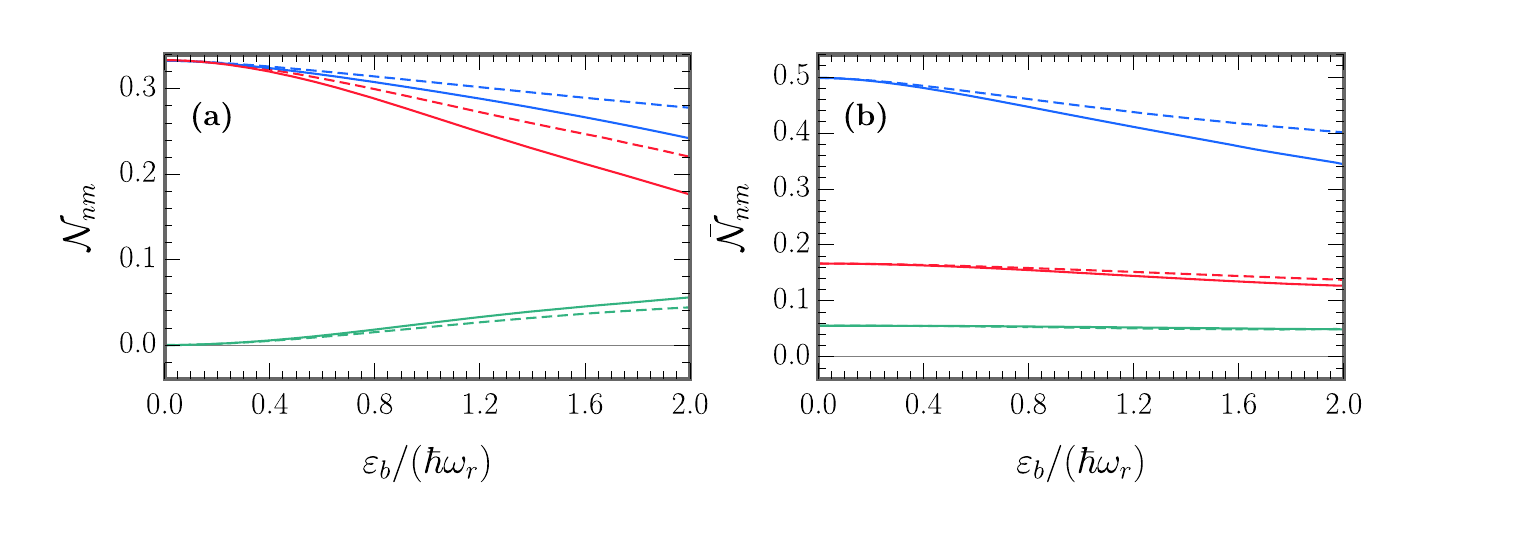}
\caption{Occupation numbers of the one-body density matrix~(a) and the reduced two-body density matrix~(b) as functions of the binding energy for the ground state of $3+3$ fermions.  Results are shown for quantum numbers of $n=0$ and $|m|=0,\,1,\,2$ [blue, red, and green lines, respectively].  Solid lines correspond to an effective range of $r_\mathrm{2D}/l_r^2=-\hspace{+0.2mm}0.2$ and dashed lines to $r_\mathrm{2D}/l_r^2=-\hspace{+0.2mm}0.001\approx0$, with the latter taken from the two lowest panels of Fig.~\ref{fig:Fig_Occ-Nums}.}
\label{fig:Fig_Occ-Nums_2}
\end{figure}

\vspace{1mm}

\subsection{Finite Effective Range}
\label{sec:Finite_Effective_Range}

Here, we examine how the effective range influences the energetic and structural properties of the $3+3$ Fermi system.  Figures~\ref{fig:Fig_ES_2}\hspace{+0.45mm}--\hspace{+0.30mm}\ref{fig:Fig_Mom-Dists_2} present our results for $r_\mathrm{2D}/l_r^2=-\hspace{+0.2mm}0.2$ --- which was the largest negative effective range considered in Ref.~\cite{Laird_2024} --- overlaid with our earlier results for very nearly zero effective range, $r_\mathrm{2D}/l_r^2=-\hspace{+0.2mm}0.001\approx0$.

\begin{figure}[H]
\vspace{-3mm}
\hspace{-5.4mm}
\includegraphics[trim = 0mm 0mm 0mm 0mm, clip, scale = 0.68]{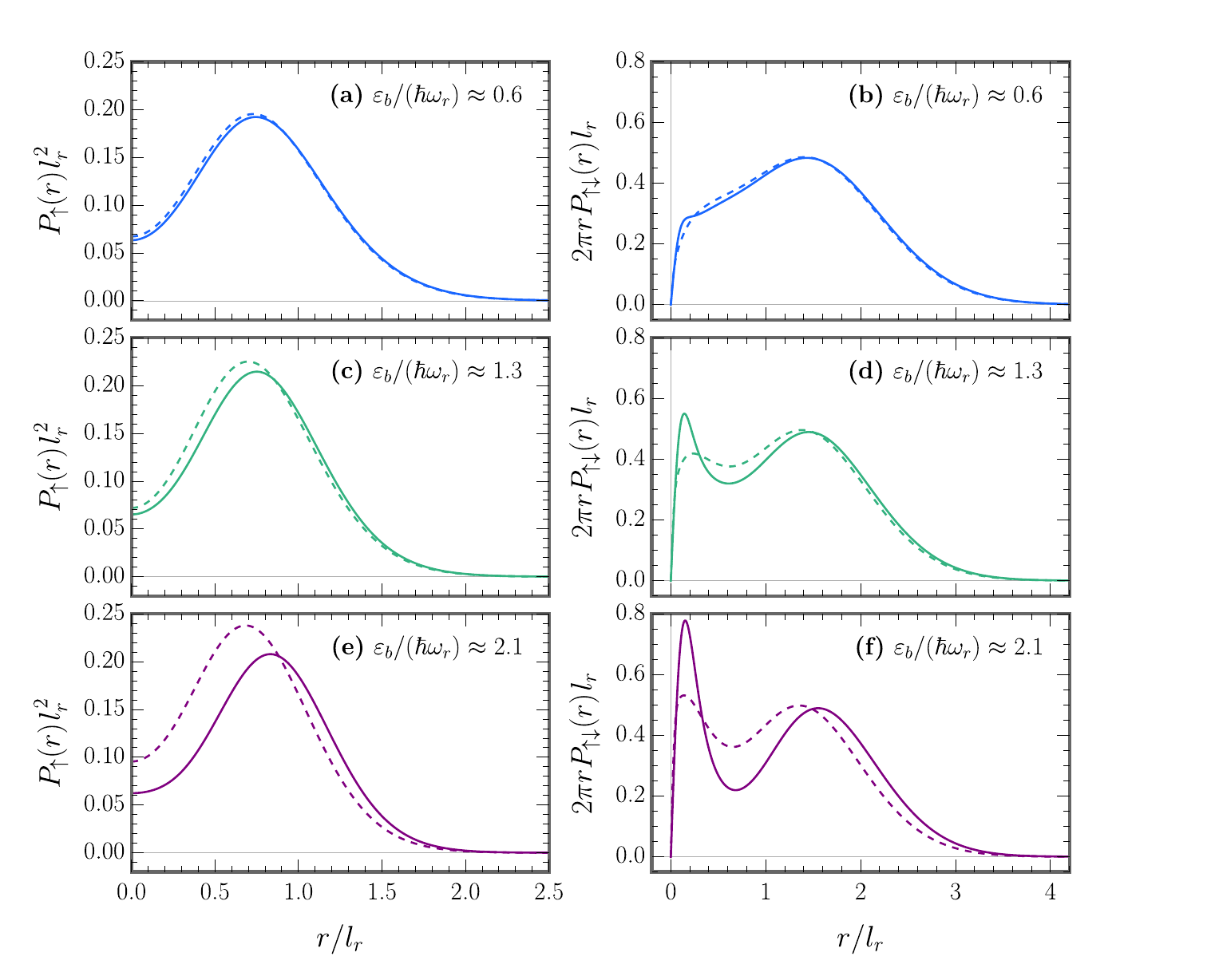}
\caption{The radial one-body density [panels (a), (c), (e)] and the (scaled) radial pair distribution function [panels (b), (d), (f)] for the $3+3$ fermion ground state at three binding energies.  Solid lines correspond to an effective range of $r_\mathrm{2D}/l_r^2=-\hspace{+0.2mm}0.2$ and dashed lines to $r_\mathrm{2D}/l_r^2=-\hspace{+0.2mm}0.001\approx0$, with the latter taken from the two lowest panels of Fig.~\ref{fig:Fig_RD_&_UDPD_new}.}
\label{fig:Fig_Rad-Dens_2}
\end{figure}

\begin{figure}[H]
%\vspace{0mm}
\hspace{-5.5mm}
\includegraphics[trim = 0mm 0mm 0mm 0mm, clip, scale = 0.679]{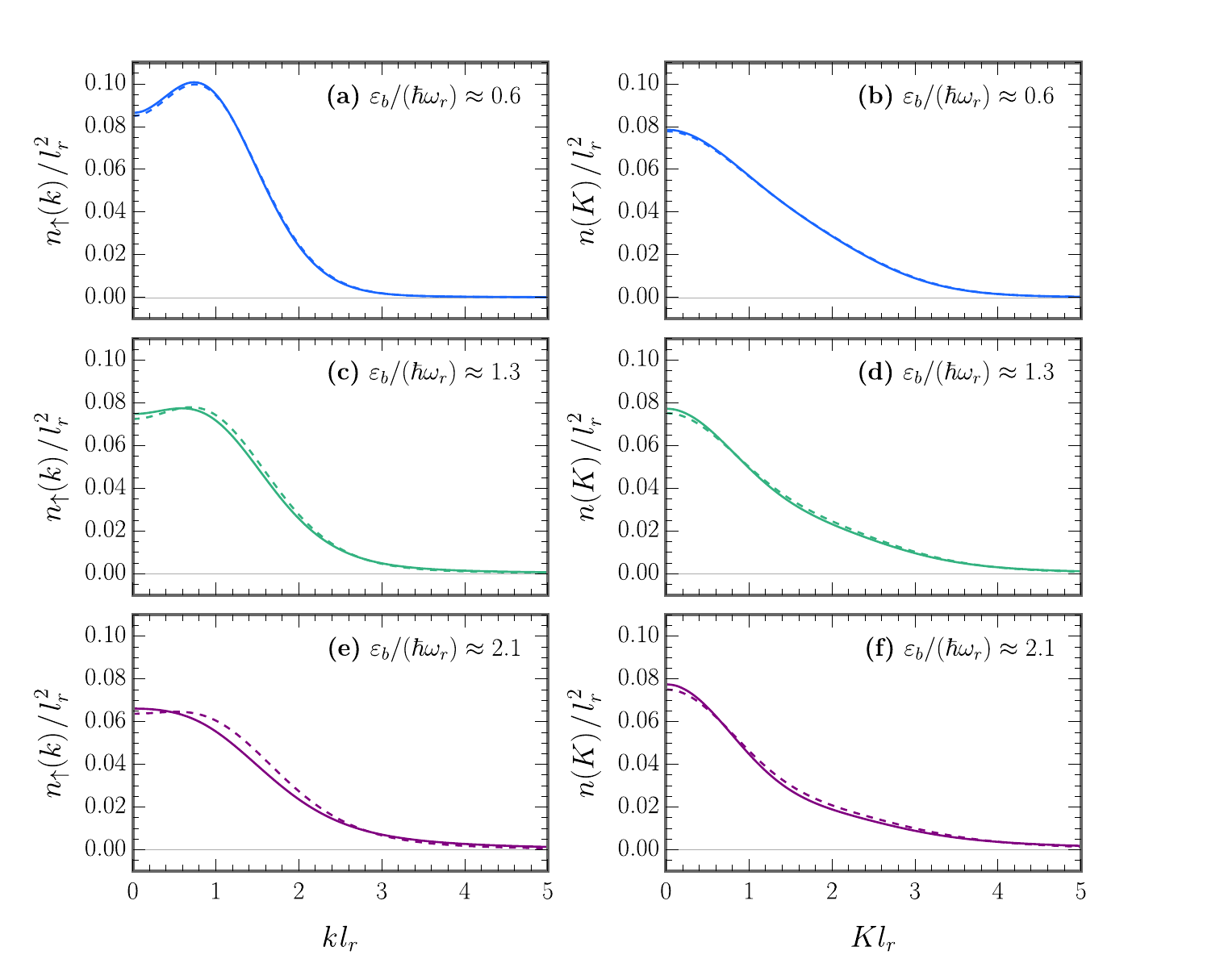}
\caption{Ground-state momentum distributions for the motion of spin-$\up$ atoms [panels (a), (c), (e)] and the centre-of-mass motion of spin-$\up$-spin-$\down$ pairs [panels (b), (d), (f)] in the $3+3$ Fermi system at three binding energies.  Solid lines correspond to an effective range of $r_\mathrm{2D}/l_r^2=-\hspace{+0.2mm}0.2$ and dashed lines to $r_\mathrm{2D}/l_r^2=-\hspace{+0.2mm}0.001\approx0$, with the latter taken from the two lowest panels of Fig.~\ref{fig:Fig_Mom_Dist}.}
\label{fig:Fig_Mom-Dists_2}
\end{figure}

Figure~\ref{fig:Fig_ES_2} shows that increasing $|r_\mathrm{2D}|$ shifts the energies upwards, with the effect more pronounced at higher binding energies.  As seen in Fig.~\ref{fig:Fig_Occ-Nums_2}, for larger $|r_\mathrm{2D}|$ the occupation numbers\linebreak of the lowest natural orbitals of the one-body density matrix decrease, implying that those of higher excited natural orbitals increase from zero.  The expansion of a tight composite
bosonic wave function over effective single-particle orbitals (the natural orbitals) generally requires significantly more terms than the expansion of an antisymmetric fermionic wave function.  This suggests that increasing $|r_\mathrm{2D}|$ drives the system further into the BEC regime.  The (scaled) radial pair distribution function displayed in Fig.~\ref{fig:Fig_Rad-Dens_2} supports this conclusion as larger $|r_\mathrm{2D}|$ increases the weight of the peak at smaller $r$, thus increasing the probability of forming a molecule. Again, these effects on the structural properties are stronger at higher binding ener-\linebreak gies.  An exception, however, is the set of momentum distributions shown in Fig.~\ref{fig:Fig_Mom-Dists_2} which re-\linebreak main essentially unaffected even at the largest considered binding energy.

\section{Conclusions}
\label{sec:Conclusion}

In summary, in this paper we have used the explicitly correlated Gaussian method to obtain a broad range of energetic and structural properties of two-dimensional trapped mesoscopic Fermi gases.  For the same range of binding energies considered in our previous work~\cite{Laird_2024}\hspace{+0.1mm}, we computed the energy spectra of the ground and low-lying excited states;  analysed non-local ground-state correlations derived from the one- and two-body density matrices; and examined local ground-state correlations using the radial and pair distribution functions.  From the den-\newpage

\vspace{5.6mm}

\noindent sity matrices we extracted not only the occupation numbers of the natural orbitals, but also the momentum distributions of atoms and pairs by means of an analytical Fourier transformation.  Additionally, we tested a measure of the molecular `\hspace{+0.1mm}condensate fraction\hspace{+0.1mm}' originally proposed for the three-dimensional case in Ref.\cite{Blume_2011}\hspace{+0.1mm}; however, its application here yielded ambiguous results. A limitation of our approach (also shared by Ref.\cite{Blume_2011}\hspace{+0.2mm}) is that in treating the two-body density matrix, the large number of degrees of freedom led us to consider only correlations between spin-$\up$-spin-$\down$ pairs characterised by the same relative-distance vector.  This means all other correlations were neglected in the calculation of quantities based on the two-body den-\linebreak sity matrix: namely, the occupation numbers $\bar{\mathcal{N}}_{nm}$\hspace{+0.2mm}, the momentum distribution $n\hspace{+0.1mm}(K)$\hspace{+0.2mm}, and the condensate fraction $\mathcal{N}_\mathrm{cond}$\hspace{+0.1mm}.  Our results consistently suggest that for up to six particles at ze-\linebreak ro effective range, binding energies of $\varepsilon_b\lesssim2\hspace{+0.2mm}\hbar\omega_r$ remain outside the regime of very strong interactions where the fermions would form tightly bound bosonic molecules.  Nonetheless, at a fixed binding energy pairing can be enhanced by introducing a finite, negative effective range --- i.e., by tuning the trap aspect ratio from strictly two-dimensional towards a more quasi-two-dimensional geometry.

\vspace{5.6mm}

While the ECG method is widely recognised for its effectiveness in solving few-body cold-atom problems~\cite{Laird_2024,Suzuki_&_Varga_1998,Review_ECG_Method}\hspace{+0.1mm}, our work has revealed two important drawbacks:  First, because tight composite bosonic wave functions are difficult to model numerically, the ECG method requires large basis sizes to converge at binding energies of $\varepsilon_b\gtrsim2\hspace{+0.2mm}\hbar\omega_r$ --- making calculations for $3+3$ or more fermions in this regime impractically slow.  Second, the principal limiting factor on computational time for $\varepsilon_b\lesssim2\hspace{+0.2mm}\hbar\omega_r$
is the number of permutations required to anti-\linebreak symmetrise the wave function, which increases factorially with the number of particles.  As a result, evaluating matrix elements for $4+4$ or more fermions at \textit{any} binding energy becomes prohibitively time-consuming, and even computing the first several excited states for $3+3$ fermions takes a significantly long time. 
This implies that the ECG method would not be well suited to performing calculations at finite temperature (where the density matrix becomes a sum over the ground and excited states, weighted by the Boltzmann factor) or to performing time dynamics (where one projects an initial wave function onto a new time-evolved basis, potentially acquiring non-zero excited-state components).

\section*{Acknowledgements}

The authors would like to thank Xiangyu (Desmond) Yin for writing the first iteration of the C code that diagonalises a given Hamiltonian via the method of explicitly correlated Gaussians.

\paragraph{Funding Information}
This research was supported by the Australian Research Council Centre of Excellence in Future Low-Energy Electronics and Technologies, also known as `\hspace{+0.25mm}FLEET\hspace{+0.50mm}'\linebreak (Project No.~CE170100039), and was funded by the Australian government.  Emma Laird re-\linebreak ceived funding from a Women--in--FLEET research fellowship.

\begin{appendix}
\numberwithin{equation}{section}

\section{Analytical Results in the Non\hspace{+0.1mm}-\hspace{+0.1mm}Interacting Limit}
\label{sec:Analytical_Results_in_the_Non-Interacting_Limit}

In this appendix we analytically derive all the occupation numbers of the projected one-body density matrix $\mathcal{N}_{nm}$ and the projected reduced two-body density matrix $\bar{\mathcal{N}}_{nm}$ for the trapped non-interacting $2+2$ atomic Fermi gas in the ground state.  In two-dimensional position space the ground-state wave function is given by
\begin{align}
\label{eq:2+2_WF}
\Psi_\mathrm{2+2}^{(\mathrm{GS})}\hspace{+0.1mm}(\r_1^\up,\,\r_2^\down,\,\r_3^\up,\,\r_4^\down)=\frac{1}{\sqrt{2}\pi^2l_r^6}\,\mathrm{exp}\hspace{-0.4mm}\left[-\sum_{i\,=\,1}^4\frac{(\r_i^{\hspace{+0.2mm}\sigma})^{\hspace{+0.2mm}2}}{2l_r^2}\right](\r_1^\up-\r_3^\up)^T(\r_2^\down-\r_4^\down)\;,
\end{align}
with $\sigma=\hspace{+0.2mm}\,\up,\,\down\hspace{+0.2mm}$.  It can readily be confirmed that Eq.~\eqref{eq:2+2_WF} is normalised and correctly results in a total ground-state energy of $E_\mathrm{com}\hspace{-0.2mm}+\hspace{+0.2mm}E_\mathrm{rel}=6\hspace{+0.3mm}\hbar\omega_r$\hspace{+0.1mm}.  As defined in Eq.~\eqref{eq:rho-up}, the corresponding one-body density matrix is
\begin{align}
[\hspace{+0.2mm}\rho_\up(\r,\,\r')]_{\hspace{+0.1mm}\mathrm{GS}}&=\int\!\cdots\!\int{d}\r_2^\down\hspace{+0.30mm}{d}\r_3^\up\hspace{+0.20mm}{d}\r_4^\down\;\Psi_\mathrm{2+2}^{(\mathrm{GS})}\hspace{+0.1mm}(\r,\,\r_2^\down,\,\r_3^\up,\,\r_4^\down)\;\Psi^{\mathrm{(GS)\,*}}_{2+2}(\r'\hspace{-0.5mm},\,\r_2^\down,\,\r_3^\up,\,\r_4^\down)\nn\\
&=\frac{1}{2\pi}\,\mathrm{exp}\hspace{+0.1mm}\!\left\{-\hspace{+0.20mm}\frac12\hspace{+0.25mm}\Big[\r^{\hspace{+0.30mm}2}+\hspace{+0.25mm}(\r')^{\hspace{+0.20mm}2}\Big]\hspace{0.0mm}\right\}\hspace{-0.4mm}(1+\r^T\hspace{-0.2mm}\r')\;.
\end{align}
Writing $\r^T\hspace{-0.2mm}\r'=r r' \mathrm{cos}\hspace{+0.4mm}(\theta-\theta')$ and then applying Eq.~\eqref{eq:partial-wave_projections} yields
\begin{subequations}
\begin{align}
&[\hspace{+0.2mm}\rho_\up^{m\;=\;0}(r,\,r')]_{\hspace{+0.1mm}\mathrm{GS}}=\mathrm{exp}\hspace{+0.1mm}\!\left\{-\hspace{+0.20mm}\frac12\hspace{+0.25mm}\Big[r^{\hspace{+0.10mm}2}+\hspace{+0.25mm}(r')^{\hspace{+0.20mm}2}\Big]\hspace{0.0mm}\right\}\hspace{-0.1mm},\\
&[\hspace{+0.2mm}\rho_\up^{m\;=\;\pm1}(r,\,r')]_{\hspace{+0.1mm}\mathrm{GS}}=\frac12\,\mathrm{exp}\hspace{+0.1mm}\!\left\{-\hspace{+0.20mm}\frac12\hspace{+0.25mm}\Big[r^{\hspace{+0.10mm}2}+\hspace{+0.25mm}(r')^{\hspace{+0.20mm}2}\Big]\hspace{0.0mm}\right\}\hspace{-0.2mm}rr'\hspace{-0.8mm},\\
&[\hspace{+0.2mm}\rho_\up^{m\;\geq\;2\hspace{+0.2mm}}(r,\,r')]_{\hspace{+0.1mm}\mathrm{GS}}=0\;\hspace{-0.1mm}.
\end{align}
\end{subequations}
Finding the eigenvalues of $\sqrt{r}\,[\hspace{+0.2mm}\rho_\up^m(r,\,r')]_{\hspace{+0.1mm}\mathrm{GS}}\sqrt{r'}\Delta{r}$ affords $\mathcal{N}_{0\hspace{+0.1mm},\,0}=1/2$ and $\mathcal{N}_{0\hspace{+0.1mm},\,\pm1}=1/4$  (with all other occupation numbers zero), consistent with the left middle panel of Fig.~\ref{fig:Fig_Occ-Nums}.

Similarly, the relevant two-body density matrix is
\begin{align}
&[\hspace{+0.2mm}\rho(\r_\up,\,\r_\up';\,\r_\down,\,\r_\down')]_{\hspace{+0.1mm}\mathrm{GS}}=\int\!\cdots\!\int{d}\r_3^\up\hspace{+0.20mm}{d}\r_4^\down\;\Psi_\mathrm{2+2}^{(\mathrm{GS})}(\r_\up,\,\r_\down,\,\r_3^\up,\,\r_4^\down)\;\Psi^\mathrm{(GS)\,*}_{2+2}(\r_\up',\,\r_\down',\,\r_3^\up,\,\r_4^\down)\nn\\
&\hspace{-0.8mm}=\frac{1}{4\pi^2}\,\mathrm{exp}\hspace{+0.1mm}\!\left\{-\hspace{+0.2mm}\frac12\hspace{+0.25mm}\Big[\r_\up^{\hspace{+0.30mm}2}+\hspace{+0.2mm}(\r_\up')^{\hspace{+0.20mm}2}+\hspace{+0.2mm}\r_\down^{\hspace{+0.30mm}2}+\hspace{+0.2mm}(\r_\down')^{\hspace{+0.20mm}2}\Big]\hspace{0.0mm}\right\}\hspace{-0.6mm}\bigg\{1+\r_\up^T\r_\up'+\r_\down^T\r_\down'+2\,\hspace{-0.1mm}(\r_\up^T\r_\down)\,\hspace{-0.2mm}\Big[(\r_\up')^T\r_\down'\Big]\bigg\}\hspace{+0.4mm},
\end{align}
as defined in Eq.~\eqref{eq:rho}.  By transforming to the centre-of-mass and relative co-ordinates of the two spin-$\up$-spin-$\down$ pairs we arrive at
\begin{align}
[\hspace{+0.2mm}\rho(\R,\,\R';\,\r,\,\r')]_{\hspace{+0.1mm}\mathrm{GS}}=&\,\,\frac{1}{32\pi^2}\,\mathrm{exp}\hspace{+0.1mm}\!\left\{-\hspace{+0.4mm}\Big[\R^{2}+\hspace{+0.2mm}(\R')^{\hspace{+0.20mm}2}\Big]-\frac14\hspace{+0.25mm}\Big[\r^{\hspace{+0.30mm}2}+\hspace{+0.2mm}(\r')^{\hspace{+0.20mm}2}\Big]\hspace{0.0mm}\right\}\hspace{-0.3mm}\times\nn\\&\hspace{+1.0mm}\bigg\{8+16\hspace{+0.3mm}\R^T\R'+\hspace{+0.2mm}4\hspace{+0.3mm}\r^T\hspace{-0.2mm}\r'+\hspace{+0.2mm}\Big(4\hspace{+0.3mm}\R^2-\r^{\hspace{+0.30mm}2}\Big)\hspace{+0.2mm}\Big[4\hspace{+0.3mm}(\R')^{\hspace{+0.20mm}2}-(\r')^{\hspace{+0.20mm}2}\Big]\bigg\}\hspace{+0.3mm}.
\end{align}
Setting $\r=\r'$ and subsequently integrating over $\r$ leads to
\begin{align}
[\hspace{+0.2mm}\rho_\mathrm{red}\hspace{+0.1mm}(\R,\,\R')]_{\hspace{+0.1mm}\mathrm{GS}}=\frac{1}{2\pi}\,\mathrm{exp}\hspace{+0.1mm}\bigg\{\hspace{-0.6mm}-\hspace{-0.3mm}\Big[\R^{2}+\hspace{+0.2mm}(\R')^{\hspace{+0.20mm}2}\Big]\hspace{+0.025mm}\bigg\}\hspace{+0.2mm}\Big[\hspace{-0.1mm}3+2\hspace{+0.3mm}\R^2\hspace{+0.3mm}(\R')^{\hspace{+0.20mm}2}-\hspace{+0.2mm}(\R-\R')^T\hspace{-0.2mm}(\R-\R')\hspace{-0.1mm}\Big]\;.
\end{align}
At this point, we can expand $(\R-\R')^T\hspace{-0.2mm}(\R-\R')=R^2+\hspace{+0.2mm}(R')^{\hspace{+0.20mm}2}-2\hspace{+0.3mm}R\hspace{+0.3mm}R'\mathrm{cos}\hspace{+0.4mm}(\phi-\phi')\hspace{+0.2mm}$  and perform partial-wave projections in analogy to  Eq.~\eqref{eq:partial-wave_projections} to find that
\begin{subequations}
\begin{align}
\label{eq:m=0}
&[\hspace{+0.2mm}\rho_\mathrm{red}^{m\;=\;0}(R,\,R')]_{\hspace{+0.1mm}\mathrm{GS}}=\mathrm{exp}\hspace{+0.1mm}\bigg\{\hspace{-0.6mm}-\hspace{-0.3mm}\Big[R^{\hspace{+0.1mm}2}+\hspace{+0.2mm}(R')^{\hspace{+0.20mm}2}\Big]\hspace{+0.025mm}\bigg\}\hspace{-0.2mm}\bigg\{3+2\hspace{+0.30mm}(R\hspace{+0.30mm}R')^{\hspace{+0.20mm}2}-\hspace{-0.3mm}\Big[R^{\hspace{+0.1mm}2}+\hspace{+0.2mm}(R')^{\hspace{+0.20mm}2}\Big]\bigg\}\hspace{+0.3mm},\\
\label{eq:m=1}
&[\hspace{+0.2mm}\rho_\mathrm{red}^{m\;=\;\pm1}(R,\,R')]_{\hspace{+0.1mm}\mathrm{GS}}=\mathrm{exp}\hspace{+0.1mm}\bigg\{\hspace{-0.6mm}-\hspace{-0.3mm}\Big[R^{\hspace{+0.1mm}2}+\hspace{+0.2mm}(R')^{\hspace{+0.20mm}2}\Big]\hspace{+0.025mm}\bigg\}\hspace{+0.4mm}R\hspace{+0.30mm}R',\\
\label{eq:m=2}
&[\hspace{+0.2mm}\rho_\mathrm{red}^{m\;\geq\;2\hspace{+0.2mm}}(R,\,R')]_{\hspace{+0.1mm}\mathrm{GS}}=0\;\hspace{-0.1mm},
\end{align}
\end{subequations}
where $\phi^{\hspace{-0.2mm}(}$$'$$^{)}$ is the angle associated with the vector $\R^{\hspace{+0.1mm}(}$$'$$^{)}\hspace{-0.4mm}$.  The occupation numbers can now be\linebreak obtained as the eigenvalues of $\sqrt{R}\,[\hspace{+0.2mm}\rho_\mathrm{red}^m\hspace{+0.1mm}(R,\,R')]_{\hspace{+0.1mm}\mathrm{GS}}\sqrt{R'}\Delta{R}$. The first of the above relations~\eqref{eq:m=0} gives $\bar{\mathcal{N}}_{0\hspace{+0.1mm},\,0}=0.625$ and $\bar{\mathcal{N}}_{1,\,0}=0.125$, and the second~\eqref{eq:m=1} gives $\bar{\mathcal{N}}_{0\hspace{+0.1mm},\,\pm1}=0.125$, while all other occupation numbers vanish --- in agreement with the right middle panel of Fig.~\ref{fig:Fig_Occ-Nums}.

\end{appendix}

\raggedright

\bibliography{SciPost_BibTeX_File.bib}

\begin{thebibliography}{10}
\providecommand{\url}[1]{\texttt{#1}}
\providecommand{\urlprefix}{URL }
\expandafter\ifx\csname urlstyle\endcsname\relax
  \providecommand{\doi}[1]{doi:\discretionary{}{}{}#1}\else
  \providecommand{\doi}{doi:\discretionary{}{}{}\begingroup \urlstyle{rm}\Url}\fi
\providecommand{\eprint}[2][]{\url{#2}}

\bibitem{Wenz_2013}
A.~N. Wenz, G.~Zürn, S.~Murmann, I.~Brouzos, T.~Lompe and S.~Jochim,
\newblock \emph{{F}rom few to many: {O}bserving the formation of a {F}ermi sea one atom at a time},
\newblock Science \textbf{342}, 457–460 (2013),
\newblock \doi{10.1126/science.1240516}.

\bibitem{Zinner_2014}
N.~T. Zinner,
\newblock \emph{{F}ew-body physics in a many-body world},
\newblock Few-Body Systems \textbf{55}, 599–604 (2014),
\newblock \doi{10.1007/s00601-014-0802-x}.

\bibitem{Grining_2015}
T.~Grining, M.~Tomza, M.~Lesiuk, M.~Przybytek, M.~Musia\l{}, R.~Moszynski, M.~Lewenstein and P.~Massignan,
\newblock \emph{{C}rossover between few and many fermions in a harmonic trap},
\newblock Physical Review A \textbf{92}, 061601 (2015),
\newblock \doi{10.1103/PhysRevA.92.061601}.

\bibitem{Rammelmuller_2016}
L.~Rammelm\"uller, W.~J. Porter and J.~E. Drut,
\newblock \emph{{G}round state of the two-dimensional attractive {F}ermi gas: {E}ssential properties from few to many body},
\newblock Physical Review A \textbf{93}, 033639 (2016),
\newblock \doi{10.1103/PhysRevA.93.033639}.

\bibitem{Levinsen_2017}
J.~Levinsen, P.~Massignan, S.~Endo and M.~M. Parish,
\newblock \emph{{U}niversality of the unitary {F}ermi gas: {A} few-body perspective},
\newblock Journal of Physics B: Atomic, Molecular and Optical Physics \textbf{50}, 072001 (2017),
\newblock \doi{10.1088/1361-6455/aa5a1e}.

\bibitem{Ran_2017}
S.-J. Ran, A.~Piga, C.~Peng, G.~Su and M.~Lewenstein,
\newblock \emph{Few-body systems capture many-body physics: Tensor network approach},
\newblock Physical Review B \textbf{96}, 155120 (2017),
\newblock \doi{10.1103/PhysRevB.96.155120}.

\bibitem{Schiulaz_2018}
M.~Schiulaz, M.~Távora and L.~F. Santos,
\newblock \emph{{F}rom few- to many-body quantum systems},
\newblock Quantum Science and Technology \textbf{3}, 044006 (2018),
\newblock \doi{10.1088/2058-9565/aad913}.

\bibitem{Bayha_2020}
L.~Bayha, M.~Holten, R.~Klemt, K.~Subramanian, J.~Bjerlin, S.~M. Reimann, G.~M. Bruun, P.~M. Preiss and S.~Jochim,
\newblock \emph{{O}bserving the emergence of a quantum phase transition shell by shell},
\newblock Nature \textbf{587}, 583–587 (2020),
\newblock \doi{10.1038/s41586-020-2936-y}.

\bibitem{Holten_2022}
M.~Holten, L.~Bayha, K.~Subramanian, S.~Brandstetter, C.~Heintze, P.~Lunt, P.~M. Preiss and S.~Jochim,
\newblock \emph{{O}bservation of {C}ooper pairs in a mesoscopic two-dimensional {F}ermi gas},
\newblock Nature \textbf{606}, 287–291 (2022),
\newblock \doi{10.1038/s41586-022-04678-1}.

\bibitem{Serwane_2011}
F.~Serwane, G.~Z\"urn, T.~Lompe, T.~B. Ottenstein, A.~N. Wenz and S.~Jochim,
\newblock \emph{{D}eterministic preparation of a tunable few-fermion system},
\newblock Science \textbf{332}, 336–338 (2011),
\newblock \doi{10.1126/science.1201351}.

\bibitem{Laird_2024}
E.~K. Laird, B.~C. Mulkerin, J.~Wang and M.~J. Davis,
\newblock \emph{{When does a Fermi puddle become a Fermi sea? Emergence of pairing in two-dimensional trapped mesoscopic Fermi gases}},
\newblock SciPost Physics \textbf{17}, 163–200 (2024),
\newblock \doi{10.21468/SciPostPhys.17.6.163}.

\bibitem{Varga_&_Suzuki_1995}
K.~Varga and Y.~Suzuki,
\newblock \emph{{P}recise solution of few-body problems with the stochastic variational method on a correlated {G}aussian basis},
\newblock Physical Review C \textbf{52}, 2885–2905 (1995),
\newblock \doi{10.1103/PhysRevC.52.2885}.

\bibitem{Varga_&_Suzuki_1996}
K.~Varga and Y.~Suzuki,
\newblock \emph{{S}tochastic variational method with a correlated {G}aussian basis},
\newblock Physical Review A \textbf{53}, 1907–1910 (1996),
\newblock \doi{10.1103/PhysRevA.53.1907}.

\bibitem{Suzuki_&_Varga_1998}
Y.~Suzuki and K.~Varga,
\newblock \emph{{S}tochastic {V}ariational {A}pproach to {Q}uantum {M}echanical {F}ew-{B}ody {P}roblems},
\newblock Springer Publishing,
\newblock ISBN 978-3-5404-9541-3 (1998).

\bibitem{Review_ECG_Method}
J.~Mitroy, S.~Bubin, W.~Horiuchi, Y.~Suzuki, L.~Adamowicz, W.~Cencek, K.~Szalewicz, J.~Komasa, D.~Blume and K.~Varga,
\newblock \emph{{T}heory and application of explicitly correlated {G}aussians},
\newblock Reviews of Modern Physics \textbf{85}, 693–749 (2013),
\newblock \doi{10.1103/RevModPhys.85.693}.

\bibitem{Boys_1960}
S.~F. Boys,
\newblock \emph{{T}he integral formulae for the variational solution of the molecular many-electron wave equation in terms of {G}aussian functions with direct electronic correlation},
\newblock Proceedings of the Royal Society of London A \textbf{258}, 402–411 (1960),
\newblock \doi{10.1098/rspa.1960.0195}.

\bibitem{Singer_1960}
K.~Singer,
\newblock \emph{{T}he use of {G}aussian (exponential quadratic) wave functions in molecular problems --- {I}. {G}eneral formulae for the evaluation of integrals},
\newblock Proceedings of the Royal Society of London A \textbf{258}, 412–420 (1960),
\newblock \doi{10.1098/rspa.1960.0196}.

\bibitem{Verhaar_1984}
B.~J. Verhaar, J.~P. H.~W. van~den Eijnde, M.~A.~J. Voermans and M.~M.~J. Schaffrath,
\newblock \emph{{S}cattering length and effective range in two dimensions: {A}pplication to adsorbed hydrogen atoms},
\newblock Journal of Physics A: Mathematical and General \textbf{17}, 595–598 (1984),
\newblock \doi{10.1088/0305-4470/17/3/020}.

\bibitem{Adhikari_1986_A}
S.~K. Adhikari,
\newblock \emph{{{Q}uantum scattering in two dimensions}},
\newblock American Journal of Physics \textbf{54}, 362–367 (1986),
\newblock \doi{10.1119/1.14623}.

\bibitem{Adhikari_1986_B}
S.~K. Adhikari, W.~G. Gibson and T.~K. Lim,
\newblock \emph{{{E}ffective‐range theory in two dimensions}},
\newblock Journal of Chemical Physics \textbf{85}, 5580–5583 (1986),
\newblock \doi{10.1063/1.451572}.

\bibitem{Levinsen_2013}
J.~Levinsen and M.~M. Parish,
\newblock \emph{{B}ound states in a quasi-two-dimensional {F}ermi gas},
\newblock Physical Review Letters \textbf{110}, 055304 (2013),
\newblock \doi{10.1103/PhysRevLett.110.055304}.

\bibitem{Kirk_2017}
T.~Kirk and M.~M. Parish,
\newblock \emph{{T}hree-body correlations in a two-dimensional $\mathrm{SU(3)}$ {F}ermi gas},
\newblock Physical Review A \textbf{96}, 053614 (2017),
\newblock \doi{10.1103/PhysRevA.96.053614}.

\bibitem{Hu_2019}
H.~Hu, B.~C. Mulkerin, U.~Toniolo, L.~He and X.-J. Liu,
\newblock \emph{{R}educed quantum anomaly in a quasi-two-dimensional {F}ermi superfluid: {S}ignificance of the confinement-induced effective range of interactions},
\newblock Physical Review Letters \textbf{122}, 070401 (2019),
\newblock \doi{10.1103/PhysRevLett.122.070401}.

\bibitem{Yin_2020}
X.~Y. Yin, H.~Hu and X.-J. Liu,
\newblock \emph{{F}ew-body perspective of a quantum anomaly in two-dimensional {F}ermi gases},
\newblock Physical Review Letters \textbf{124}, 013401 (2020),
\newblock \doi{10.1103/PhysRevLett.124.013401}.

\bibitem{Blume_2011}
D.~Blume and K.~Daily,
\newblock \emph{{T}rapped two-component {F}ermi gases with up to six particles: {E}nergetics, structural properties, and molecular condensate fraction},
\newblock Comptes Rendus Physique \textbf{12}, 86–109 (2011),
\newblock \doi{10.1016/j.crhy.2010.11.010}.

\bibitem{von_Stecher_2008}
J.~von Stecher, C.~H. Greene and D.~Blume,
\newblock \emph{{E}nergetics and structural properties of trapped two-component {F}ermi gases},
\newblock Physical Review A \textbf{77}, 043619 (2008),
\newblock \doi{10.1103/PhysRevA.77.043619}.

\bibitem{Daily_2010}
K.~M. Daily and D.~Blume,
\newblock \emph{{E}nergy spectrum of harmonically trapped two-component {F}ermi gases: {T}hree- and four-particle problem},
\newblock Physical Review A \textbf{81}, 053615 (2010),
\newblock \doi{10.1103/PhysRevA.81.053615}.

\bibitem{Bradly_2014}
C.~J. Bradly, B.~C. Mulkerin, A.~M. Martin and H.~M. Quiney,
\newblock \emph{{C}oupled-pair approach for strongly interacting trapped fermionic atoms},
\newblock Physical Review A \textbf{90}, 023626 (2014),
\newblock \doi{10.1103/PhysRevA.90.023626}.

\bibitem{Yin_2015}
X.~Y. Yin and D.~Blume,
\newblock \emph{{T}rapped unitary two-component {F}ermi gases with up to ten particles},
\newblock Physical Review A \textbf{92}, 013608 (2015),
\newblock \doi{10.1103/PhysRevA.92.013608}.

\bibitem{Lewart_1988}
D.~S. Lewart, V.~R. Pandharipande and S.~C. Pieper,
\newblock \emph{Single-particle orbitals in liquid-helium drops},
\newblock Physical Review B \textbf{37}, 4950–4964 (1988),
\newblock \doi{10.1103/PhysRevB.37.4950}.

\bibitem{Sakurai_2010}
J.~J. Sakurai and J.~Napolitano,
\newblock \emph{{M}odern {Q}uantum {M}echanics},
\newblock Addison--Wesley Publishing, 2nd Edition,
\newblock ISBN 978-0-8053-8291-4 (2010).

\bibitem{Levinsen_&_Parish_2015}
J.~Levinsen and M.~M. Parish,
\newblock \emph{{C}hapter 1: {S}trongly interacting two-dimensional {F}ermi gases},
\newblock Annual Review of Cold Atoms and Molecules \textbf{3}, 1–75 (2015),
\newblock \doi{10.1142/9789814667746_0001}.

\bibitem{Zwerger_2012}
W.~Zwerger, ed.,
\newblock \emph{{L}ecture {N}otes in {P}hysics (vol. 836): {T}he {BCS}--{BEC} {C}rossover and the {U}nitary {F}ermi {G}as},
\newblock Springer Publishing,
\newblock ISBN 978-3-642-21977-1 (2012).

\bibitem{Braaten_2006}
E.~Braaten and H.-W. Hammer,
\newblock \emph{{U}niversality in few-body systems with large scattering length},
\newblock Physics Reports \textbf{428}, 259–390 (2006),
\newblock \doi{10.1016/j.physrep.2006.03.001}.

\bibitem{Busch_1998}
T.~Busch, B.-G. Englert, K.~Rzażewski and M.~Wilkens,
\newblock \emph{Two cold atoms in a harmonic trap},
\newblock Foundations of Physics \textbf{28}, 549–559 (1998),
\newblock \doi{10.1023/A:1018705520999}.

\bibitem{Efimov_1971}
V.~Efimov,
\newblock \emph{{W}eakly bound states of three resonantly interacting particles},
\newblock Soviet Journal of Nuclear Physics \textbf{12}, 589–601 (1971).

\bibitem{Liu_2010}
X.-J. Liu, H.~Hu and P.~D. Drummond,
\newblock \emph{Exact few-body results for strongly correlated quantum gases in two dimensions},
\newblock Physical Review B \textbf{82}, 054524 (2010),
\newblock \doi{10.1103/PhysRevB.82.054524}.

\bibitem{Bjerlin_2016}
J.~Bjerlin, S.~M. Reimann and G.~M. Bruun,
\newblock \emph{{F}ew-body precursor of the {H}iggs mode in a {F}ermi gas},
\newblock Physical Review Letters \textbf{116}, 155302 (2016),
\newblock \doi{10.1103/PhysRevLett.116.155302}.

\bibitem{Astrakharchik_2005}
G.~E. Astrakharchik, J.~Boronat, J.~Casulleras and S.~Giorgini,
\newblock \emph{{M}omentum distribution and condensate fraction of a fermion gas in the {BCS--BEC} crossover},
\newblock Physical Review Letters \textbf{95}, 230405 (2005),
\newblock \doi{10.1103/PhysRevLett.95.230405}.

\bibitem{Yan_2015}
Y.~Yan and D.~Blume,
\newblock \emph{{I}ncorporating exact two-body propagators for zero-range interactions into $\mathrm{N}$-body {M}onte {C}arlo simulations},
\newblock Physical Review A \textbf{91}, 043607 (2015),
\newblock \doi{10.1103/PhysRevA.91.043607}.

\end{thebibliography}

\end{document}